\documentclass[twocolumn]{aastex631}

\def\gsim{\mathrel{\rlap{\lower 4pt \hbox{\hskip 1pt $\sim$}}\raise 1pt
\hbox {$>$}}}
\def\lsim{\mathrel{\rlap{\lower 4pt \hbox{\hskip 1pt $\sim$}}\raise 1pt
\hbox {$<$}}}

\received{October 11, 2022}
\revised{July 29, 2023}
\accepted{July 31, 2023}

\shorttitle{Very Early-Phase Spectra of SNe Ia}
\shortauthors{Ogawa et al.}

\graphicspath{{./}{figures/}}

\usepackage{color}
\usepackage{comment}

\begin{document}

\title{Systematic Investigation of Very Early-Phase Spectra of Type Ia Supernovae}

\correspondingauthor{Mao Ogawa and Keiichi Maeda}
\email{mao.ogawa@kusastro.kyoto-u.ac.jp; keiichi.maeda@kusastro.kyoto-u.ac.jp}

\author[0000-0001-5822-1672]{Mao Ogawa}
\affiliation{Department of Astronomy, Kyoto University,
Kitashirakawa-Oiwake-Cho,
Sakyo-ku, Kyoto, 606-8502, Japan}

\author[0000-0003-2611-7269]{Keiichi Maeda}
\affiliation{Department of Astronomy, Kyoto University, 
Kitashirakawa-Oiwake-Cho, 
Sakyo-ku, Kyoto, 606-8502, Japan}

\author[0000-0002-4540-4928]{Miho Kawabata}
\affiliation{Nishi-Harima Astronomical Observatory, Center for Astronomy,
University of Hyogo, 407-2 Nishigaichi, Sayo-cho, Sayo, Hyogo, 679-5313, Japan}

\begin{abstract}
It has been widely accepted that Type Ia supernovae (SNe Ia) are thermonuclear explosions of a CO white dwarf. However, the natures of the progenitor system(s) and explosion mechanism(s) are still unclarified. Thanks to the recent development of transient observations, they are now frequently discovered shortly after the explosion, followed by rapid spectroscopic observations. In this study, by modeling very early-phase spectra of SNe Ia, we try to constrain the explosion models of SNe Ia. By using the Monte Carlo radiation transfer code, TARDIS, we estimate the properties of their outermost ejecta. We find that the photospheric velocity of normal-velocity supernovae (NV SNe) in the first week is $\sim$15000 km s$^{-1}$. The outer velocity, to which the carbon burning extends, spans the range between $\sim$20000 and 25000 km s$^{-1}$. The ejecta density of NV SNe also shows a large diversity. 
For high-velocity supernovae (HV SNe) and 1999aa-like SNe, the photospheric velocity is higher, $\sim$20000 km s$^{-1}$. They are different in the photospheric density, with HV SNe having higher density than 1999aa-like SNe. For all these types, we show that the outermost composition is closely related to the outermost ejecta density; the carbon burning layer and the unburnt carbon layer are found in the higher-density and lower-density objects, respectively. This finding suggests that there might be two sequences, the high-density and carbon-poor group (HV SNe and some NV SNe) and the low-density and carbon-rich group (1999aa-like and other NV SNe), which may be associated with different progenitor channels.
\end{abstract}

\keywords{Type Ia supernova --- Radiative transfer simulations --- White dwarfs}

\section{Introduction} \label{sec:intro}
It has been widely accepted that Type Ia supernovae (SNe Ia) are thermonuclear explosions of a massive CO white dwarf (WD) in a binary system. They show a correlation between the peak luminosity and the light-curve timescale \citep{Phillips1993}, with which SNe Ia can be used as reliable standardisable candles for the cosmological measurement \citep{Riess1998,Perlmutter1999}. 

However, observational properties of SNe Ia are not uniform. Various sequences or sub-classes have been identified \citep[e.g.,][]{Branch2006}. Classically, SNe Ia are placed on the peak-luminosity sequence, i.e., the 1991T-normal-1991bg sequence \citep{Filippenko1992a,Filippenko1992b}; 1991T-like SNe are brighter than $-19.5$ mag at the peak, while 1991bg-like SNe can be fainter than $-18.0$ mag. 
They also form a spectral sequence, associated with the photospheric temperature \citep{Nugent1995}. An additional diversity in the spectral properties within the normal class has also been discovered.  Based on the velocity of \ion{Si}{2} 6355 $\rm{\AA}$ at the maximum-light phase, the normal class can be further divided into two types \citep{Benetti2005,Wang2009}; high-velocity (HV) SNe (with $\gsim 12,000$ km s$^{-1}$) and normal-velocity (NV) SNe (with $\lsim 12,000$ km s$^{-1}$). The origins of these diversities have not yet been clarified. 

The natures of the progenitor(s) and explosion mechanism(s) of SNe Ia are still unclarified. There are two popular scenarios for their progenitors; (1) the single degenerate (SD) scenario in which the WD mass approaches nearly to the Chandrasekhar mass ($M_{ch}$) by accreting hydrogen or helium via binary mass transfer \citep{WhelanAndIben1973}, and (2) the double degenerate (DD) scenario in which merging two sub-$M_{ch}$ WDs leads to the thermonuclear runaway \citep{IbenAndTutukov1984}.

Regarding explosion mechanisms, several models have been proposed.
Classically, the delayed-detonation model of a $M_{ch}$ WD has been considered as a plausible model. In this model, carbon ignition occurs near the center of the progenitor, and the associated flame speed changes from subsonic (i.e., deflagration) to supersonic (detonation).
A number of theoretical simulations have been performed \citep[e.g.,][]{Khokhlov1991,HoeflichAndKhokhlov1996,Iwamoto1999,Roepke2007,Maeda2010_2d_ddt}.
As a global structure, the iron-group elements such as $\rm{^{56}Ni}$ and $\rm{^{58}Ni}$ are synthesized in the central region, while the lighter elements are distributed toward the outer region. The outermost region mostly consists of $\rm{^{16}O}$ due to the carbon burning, with little unburnt carbon; in the reference model CS15DD2 of \citet{Iwamoto1999}, $\sim 5\times10^{-3} \rm{M_{\odot}}$ of the unburnt carbon is distributed at $>$30,000 km s$^{-1}$. 
Since the turbulence during deflagration creates a seed for the mixing structure \citep{Seitenzahl2013}, the onion-like structure described above is smoothed out to some extent in multi-dimensional simulations.
In this model, the central density of the WD, the composition structure, and the condition of the deflagration-to-detonation transition (DDT) can potentially produce the diversity in SNe Ia \citep[e.g.,][]{Umeda1999,Iwamoto1999,HoeflichAndKhokhlov1996}. 
Moreover, since the central ignition is not necessarily spherically symmetric, a global asymmetry can be produced, which might contribute to observational diversity \citep{Maeda2010}.

Another leading model is the sub-$M_{ch}$ WD double detonation model, which has been under intensive investigation especially in the last decade. In this model, the He detonation occurs first in the He shell on the surface of a sub-$M_{ch}$ CO WD, and the resulting shock wave triggers the carbon detonation near the center of the progenitor WD \citep[e.g.,][]{Nomoto1982,HoeflichAndKhokhlov1996,Sim2010,Shen2021}. Since the original suggestion \citep[e.g.,][]{Nomoto1982}, the scenario had been relatively unexplored for several decades, as compared to the delayed-detonation model, mainly due to some shortcomings in the model to reproduce observational properties of normal SNe Ia \citep[e.g.,][]{HoeflichAndKhokhlov1996}. However, the scenario has been revisited both theoretically \citep{Fink2010,Shen2018,iwata2022} and observationally \citep{Jiang2017,maeda2018,De2019} in the last decade. The double-detonation model has been theoretically investigated both by one-dimensional simulations \citep{Sim2010,WoosleyAndKasen2011} and by multi-dimensional simulations \citep{Fink2010,Taniwaki2018,Shen2021}. 
In this model, the central region is dominated by $^{56}$Ni as the carbon detonation ash. The model has characteristic nucleosynthesis products in the outermost layer as a result of the He detonation; a large amount of unburnt He is left, with a small amount of intermediate-mass elements such as Si and Ca. Depending on the model details, the Fe-peak elements are also produced in the outermost layer. In the model by \citet{Shen2021}, oxygen and unburnt carbon are distributed at the 15000-20000 $\rm{km s^{-1}}$ near the boundary between the He shell and CO core material. The variation in the CO core mass and He shell mass is expected in this scenario; the latter is dependent either on the mass accretion rate or the companion WD mass.

Spectroscopic observation is one of the most powerful methods to understand the nature of SNe. Thanks to the recent development of transient observations, there have been an increasing number of SNe Ia discovered shortly after the explosion and quickly followed by spectroscopic observations. Given that the photosphere generally recedes in the mass coordinate due to the expansion and the density decrease, the very early-phase spectra trace the nature of the outermost layer \citep[e.g.,][]{kawabata2020}; this is a place where we may begin to decode differences associated with different explosion models, e.g., the carbon content (see above). However, systematic investigation of the very early-phase spectra through spectral modeling has been lacking, since these very-early phase spectroscopy has become feasible only recently. In addition, spectral modeling of SNe Ia has been frequently conducted for detailed time sequence of individual objects \citep[i.e., the `tomography';][]{Mazzali2014},  the strategy which has been applied to a number of objects \citep[e.g.,][for recent examples]{magee2017,heringer2019,barna2021,aouad2022,obrien2023}, but systematic investigation of a sample of spectra at similar epochs has been limited. The latter is our strategy in this work; by modeling spectra of SNe in the earliest phase ever possible and in a systematic manner, we aim at investigating the nature of the outermost layer, its diversity and a possible relation to different progenitor and explosion scenarios.

Another issue related to the SN spectral synthesis study is treatment of subjectivity and uncertainty in fitting the observational data, given a large degree of freedom involved in the spectral modeling. With recent advance in the computation power as coupled with a sophisticated method like a machine learning, a possibility of an automated fitting procedure without involving human inspection starts attracting attention; such an approach has, for example, been realized by construction of a synthetic-spectra emulator \citep{kerzendorf2021}; it has been successfully applied to fit an observed spectrum of a single object \citep{obraien2021}, or recently to a spectral sequence of a group of objects \citep{obrien2023}. In the present work, we seeks for a complementary approach between the classical method (based on experts' experience and visual inspection) and the modern method (based on automation).

In this study, we perform spectral synthesis calculations for very early-phase spectra, taken within a week since the explosion, for 14 SNe Ia. In Section \ref{sec:method}, we summarize the properties of the sample of SNe Ia we compiled from the literature, and describe the spectral fitting method. We present the results of the spectral synthesis calculations and the comparison to the observed spectra in Section \ref{sec:result}. In Section \ref{sec:implication}, we  discuss the nature of SNe Ia obtained through the density and composition structures, including discussion on emerging subclasses and their possible connections to different progenitor/explosion mechanisms. 
Section \ref{sec:discussion} provides additional discussion. Our findings are summarized in Section \ref{sec:conclusion}.

\section{Method}\label{sec:method}

\subsection{The sample selection and properties of 14 SNe Ia}
The sample of very early-phase spectra of 14 SNe Ia has been constructed as follows. We here searched for objects that satisfy the following three criteria through the astrophysics data system (ADS)\footnote{\url{https://ui.adsabs.harvard.edu/} .}; (1) spectra observed within 1 week since the explosion exist, (2) photometric data are available that cover the date when the spectrum was taken, and (3) it has been intensively observed around the maximum light, so that basic properties are well specified (e.g., peak luminosity, SN Ia classification and subtype, declining rate). We summarize the properties of these 14 SNe Ia in Table \ref{tab:object}.

\begin{deluxetable*}{lcccccccr}
\tablecaption{The properties of 14 SNe Ia studied in the present work\label{tab:object}}
\tablewidth{0pt}
\tablehead{
\colhead{Object} & \colhead{Type} & \colhead{z} & \colhead{$\mu$}& 
\colhead{$E(B-V)_{\rm{host}}$} & \colhead{$R_{V}$} & \colhead{$E(B-V)_{\rm{MW}}$} &
\colhead{$\Delta m_{15(B)}$} & \colhead{Reference}  \\
\colhead{} & \colhead{} & \colhead{} & \colhead{(mag)} &
\colhead{(mag)} &  \colhead{} & \colhead{(mag)} &
\colhead{(mag)} & \colhead{}
}
\startdata
2009ig	&	HV	&	0.00877	&	32.6	&	0.00625	&	1.6	&	0.089	&	0.89	&	\citet{Foley2012}	\\
2011fe	&	NV	&	0.000804	&	29.04	&	0.032	&	3.1	&	0.008	&	1.18	&	\citet{Zhan2016}	\\
2012cg	&	99aa	&	0.00146	&	30.9	&	0.18	&	$2.4^{(a)}$	&	0.018	&	0.86	&	\citet{Silverman2012}	\\
2012fr	&	NV	&	0.004	&	31.27	&	0.03	&	3.1	&	0.018	&	$0.83^{(b)}$	&	\citet{Contreras2018}	\\
2012ht	&	NV	&	0.004	&	31.5	&	0	&	0	&	0.02	&	1.39	&	\citet{Yamanaka2014}	\\
2013dy	&	NV	&	0.00389	&	31.49	&	0.15	&	$3.1^{(c)}$	&	0.15	&	0.89	&	\citet{Zheng2014}	\\
2013gy	&	NV	&	0.0140	&	33.68	&	0.106	&	$2.4^{(a)}$	&	0.049	&	1.234	&	\citet{Holmbo2019}	\\
2016coj	&	NV	&	0.005	&	31.9	&	0	&	3.1	&	0.02	&	1.25	&	\citet{Zeng2017}	\\
2017cbv	&	99aa	&	0.00340	&	31.14	&	0	&	0	&	0.15	&	1.06	&	\citet{Hosseinzadeh2017}	\\
2017cfd	&	NV	&	0.0121	&	33.52	&	$0.1^{(d)}$	&	1.7	&	0.02	&	1.16	&	\citet{Han2020}	\\
2017fgc	&	HV	&	0.008	&	32.81	&	0.29	&	1.55	&	0.029	&	1.1	&	\citet{Burgaz2021}	\\
2018gv	&	NV	&	0.0053	&	$30.92^{(e)}$	&	0.028	&	3.1	&	0.051	&	0.96	&	\citet{Yang2020}	\\
2019ein	&	HV	&	0.00776	&	32.95	&	0.09	&	1.55	&	0.011	&	1.36	&	\citet{Pellegrino2020}	\\
2019yvq	&	HV	&	0.00908	&	33.14	&	0.032	&	2.4	&	0.018	&	1.5	&	\citet{Miller2020}	\\
\enddata
\tablecomments{
The data come from the reference papers (9th column), unless mentioned below. 
(a) Unless the value of $R_{V}$ is discussed in the original reference, we adopt $R_{V} = 2.4$ \citep[see][]{Wang2009}.
(b) \citet{Cain2018}.
(c) \citet{Pan2015}.
(d) From Fig.5 of the reference paper, we adopt $E(B-V)_{\rm{host}}$ = 0.1.
(e) \citet{Nasonova2011}.
}
\end{deluxetable*}

Table \ref{tab:spectra} summarizes the sources of the spectra modeled in the present study\footnote{The spectra are downloaded from the WISeRep \citep{yaron2012}; \url{https://www.wiserep.org/} .}.
We first calibrate the observed flux with the photometry data. The extinction is then corrected for with the extinction law by \citet{Cardelli1989}; it is performed for the Milkey Way (with $R_{V} = 3.1$) and the host galaxy separately (see Table \ref{tab:object}). Finally, we convert the flux to the luminosity, using the distance modulus listed in Table \ref{tab:object}. In addition, the wavelength is converted to the rest-frame wavelength.

\begin{deluxetable}{llcr}
\tablecaption{The spactra modeled in the present work\label{tab:spectra}}
\tablewidth{0pt}
\tablehead{
\colhead{Object} & \colhead{Date} & \colhead{$\rm{Phase^{*}}$} &  \colhead{Reference}\\ 
\colhead{} & \colhead{} & \colhead{(day)} & \colhead{} 
}
\startdata
2009ig	&	2009-08-22	&	-14.2	&	\citet{Foley2012}	\\
2011fe	&	2011-08-25	&	-16	    &	\citet{Zhan2016}	\\
2012cg	&	2012-05-18	&	-14.8	&	\citet{Silverman2012}	\\
2012fr	&	2012-10-28	&	-14.51	&	\citet{Childress2013}	\\
2012ht	&	2012-12-20	&	-13.9	&	\citet{Yamanaka2014}	\\
2013dy	&	2013-07-11	&	-16.07	&	\citet{Zheng2014}	\\
2013gy	&	2013-12-07	&	-14.7	&	\citet{Holmbo2019}	\\
2016coj	&	2019-05-28	&	-11	    &	\citet{Zeng2017}	\\
2017cbv	&	2017-03-10	&	-19	    &	\citet{Hosseinzadeh2017}	\\
2017cfd	&	2017-03-18	&	-13.2	&	\citet{Han2020}	\\
2017fgc	&	2017-07-11	&	-13	    &	\citet{Burgaz2021}	\\
2018gv	&	2018-01-16	&	-15.5	&   \citet{2018TNSCR2076....1B} \\
2019ein	&	2019-05-02	&	-14	    &	\citet{Pellegrino2020}	\\
2019yvq	&	2020-01-01	&	-12.94	&	\citet{Burke2021}	\\
\enddata
\tablecomments{*Phase is the day measured from the maximum light. 
}
\end{deluxetable}

\subsection{Spectral Synthesis Calculation: TARDIS} \label{sec:TARDIS}
To constrain the structure of the outermost ejecta of SNe Ia, we use a radiation transfer code, TARDIS \citep{Kerzendorf2014}. For spherically symmetric and homologously expanding ejecta, it performs a radiative transport calculation by using the Monte Carlo method for photon packets emitted originally from a sharply-defined photosphere. The input parameters are as follows; (1) the luminosity, (2) the photospheric velocity, (3) the density structure, (4) the abundance structure (after explosive nucleosynthesis), and (5) the time since the explosion. Following the photon propagation, the structures of the temperature and the excitation state of the elements are constructed so as to satisfy the thermal balance as coupled with the determination of the ionization state through the ionization balance. Finally, the emerging spectrum is calculated.

We performed the spectral synthesis calculations for various combinations of the input parameters to fit the observed spectrum. Our main goal is to constrain the density and the composition distributions in the outermost ejecta. As a general problem, it involves a large degree of freedom. In the present work, in order to simplify the problem, we adopt a single power-law density structure, $\rho \propto v^{-\alpha} (\alpha=4 \sim 14)$, up to the outermost velocity (which is taken as a parameter together with $\alpha$).
 We note that we have to introduce the outermost velocity cut in the density structure, since the single power-law distribution alone does not take into account the sharp drop of the density in the outermost layer expected in (any) explosion models that connects the ejecta and the surrounding `vacuum'. For example, the W7 model \citep{nomoto1984} has the sharp drop in the density distribution at $\sim 20,000$ km s$^{-1}$, above which there is essentially no ejecta material while the single-power law provides a reasonable approximation below it. The outermost velocity is usually not important in modeling the maximum-phase spectra, given that the photosphere has already receded deep inside; this is not the case if one starts studying the infant-phase spectra, aiming at investigating the outermost density structure which is otherwise not accessible. 

We further assume a uniform composition structure above the photosphere. In addition, based on nuclear burning physics, we consider only two characteristic layers: a carbon burning layer and an unburnt C+O layer. In the carbon burning layer, the mass fractions are set as follows; 0.6795(O), 0.15(Si), 0.09(Mg), 0.07(S), 0.01(Ar), and 0.0005(Ca). In the unburnt C+O layer, we adopt the following fractions; 0.475(C), 0.5(O), and 0.025(Ne). 
Since the photosphere is near the surface of the ejecta in the very early phase, it is enough to consider these two outermost layers.  
Using these two layers, we mix the unburnt carbon layer and the carbon burning layer in proportions of $x$ and $(1-x)$, respectively. Then, we change the mixing fraction, $x$ (between 0 and 1), and make an uniform composition. Finally, we add the solar abundance to the elements with the atomic number $N=12$, $14$, $16$, and $20-30$. 

 Modeling a given spectrum is separated into two stages. As the first step,
a search for a `rough' solution is performed in a manner similar to the classical approach with visual inspection. This is done by performing TARDIS simulations for (relatively coarse) grids of models in a large parameter space. The fitting results are checked through our fitting/comparison algorithms and visual inspection (see Section 2.3). Then we exclude the parameter space that unlikely produce a good match to the observed spectrum, and set new (finer) grids of models for the final spectral synthesis simulations; the number of final model grids are typically $\sim 2000 - 3000$ for each spectrum, which allows the investigation with relatively moderate computational power.

 The first step also serves for fixing the values for the time since the explosion and luminosity in the TARDIS input parameters. For the time since the explosion, we first try to constrain the explosion date using the light curve evolution. We assume the homologously expanding `fireball model' \citep{Arnett1982}. We use the $R$-band magnitude (if not available, the $r$-band magnitude is used instead), and  fit the flux $(f)$ by $f\propto t^{2}$, where $t$ is time since the explosion. In this light curve fit, we use only the data points before $-7.0$ days measured from the maximum light; this is a reasonable compromise between the assumption of the single power law and the sufficient number of the data points. The light-curve fit result is used as an initial guess for the time since the explosion for the spectrum under consideration.  
 However, the date estimated by the light-curve evolution sometimes shows discrepancy to the `spectral' phase \citep{Mazzali2014}. We thus allow to change it with the increment of 0.5 days in the first step, to check if the spectral-fit quality is improved. A similar procedure is performed for the input luminosity. Through this process, we thus fix the time since the explosion and the luminosity, which are adopted in setting up the final model grids.

\subsection{Fitting procedure; numerical evaluation of the quality of the fit}\label{sec:analysis}
In many previous works for SN spectral synthesis, the quality of the spectral fit was checked and the best-fit parameters were selected by visual inspection \citep[e.g.,][]{Mazzali2014,Kwok2022}. 
However, since the spectral synthesis calculation involves many parameters even for a simplified/idealized input model, we have decided to introduce an objectively-quantified index to justify the spectral fit and check possible degeneracy between the parameters.

A powerful approach to fully replace the classical (human-based) method is the development of a spectral-synthesis emulator \citep{kerzendorf2021} and the direct data-cube comparison between models and an observed spectrum \citep{obraien2021}. If one wants to select a best-fit synthesis spectrum with a number of model parameters, one will need a huge number of spectral synthesis simulations. This can be reduced to the practically manageable number by introducing a machine learning, i.e., emulator. This is an approach adopted by \citet{obraien2021} to model a single spectrum of SN 2020bo. This is a powerful method, but for the present purpose it is useful to consider another, complementary approach: (1) To model the number of observed spectra which cover a range of epochs and spectral features, a training set will need to be carefully constructed and investigated, which may require a huge number of spectral simulation calculations to train the emulator for practical application \citep[but see,][for recent update along this line]{obrien2023}. (2) While the application of the emulator to the direct data-cube fit to the data has been successful to obtain overall properties, it has not been clarified whether such a method is sufficiently sensitive to pick up relatively weak features in the fit (such as S II and C II) \citep[see, e.g., Figure 2 of][]{obraien2021}; it is indeed a main interest in the present investigation.

 We thus introduce our own-customized method for the fit of the model spectra to the observational spectra, based on the `pre-extracted features' that characterize the properties of SN spectra (which relies on the experiences of experts and the experiment by our own; see below). Rather than directly comparing the model and observed spectra both as the one-dimensional data set with thousands of data points as a function of the wavelength, we have decided to extract some characteristics spectral features as motivated by the physics of spectral formation. The first item then is to decide which `observational' features are to be fit. We have tested various observational quantities as follows; first we have evaluated how well an observed spectrum matches to a TARDIS model spectrum, solely based on the fit between the characteristic quantities under consideration, which are extracted both from the observed spectrum and the TARDIS model using the same scheme. The result is then checked by visual inspection of the match directly between the observed and model spectra. By repeating these procedures, we have decided to adopt the following items; (a) the velocity minimum ($\lambda_{\rm{min}}$) of each absorption line, (b) the Equivalent width (EW) of each absorption line, (c) the full-width half maximum (FWHM) of each absorption line, and (d) the color of the spectrum (Col). The absorption lines adopted for the tests (a)-(c) are as follows; (1) \ion{Fe}{2} \& \ion{Fe}{3} 4500\rm{\AA}, (2) \ion{Fe}{2} \& \ion{Fe}{3} 5200\rm{\AA}, (3) \ion{S}{2} 5454 \rm{\AA}, (4) \ion{S}{2} 5620 \rm{\AA}, (5) \ion{Si}{2} 5972 \rm{\AA}, (6) \ion{Si}{2} 6355 \rm{\AA}, and (7) \ion{C}{2} 6578 \rm{\AA}.

For an observational spectrum, the properties of the lines (e.g., EWs) are extracted as follows. 
First, we provide a guess to the blue and red edges of each absorption line in the wavelength by visual inspection. 
We then numerically search for the wavelength at which the flux takes the maximum value (i.e., the true edge) within 30 $\rm{\AA}$ around the initial guess, for both blue and red sides. 
The imposed range of 30\rm{\AA} is sufficient as it corresponds to $\sim  1,500$ km s$^{-1}$, which is larger than the accuracy of the initial guess through the visual inspection. 

A continuum is set as a straight line connecting the blue and red edges of each absorption line,  following the standard procedure to compute the `pseudo' EWs \citep[e.g.,][]{Hachinger2008,Zhao2015,modjaz2016}. 
Then, a normalized line profile is constructed by using the continuum for each absorption line, which is finally used to define (a) $\lambda_{\rm{minobs}}(i)$ (the velocity at the absorption minimum), (b) $\rm{EW_{obs}}(i)$ (the equivalent width), and (c) $\rm{FWHM_{obs}}(i)$  (the line width), where the index $i$ (from 1 through 7) refers to the different absorption lines (see above). 
For the color (item d), denoted as $\rm{Col_{obs}}$, we calculate the slope of the line connecting the flux at the red edge of the \ion{Fe}{2} \& \ion{Fe}{3} 5200 \rm{\AA} feature and the flux at 6700 \rm{\AA}. 
Note that the detail in the determination of the continuum (a strait-line continuum in this case) is not a concern; the key is that the same procedure is applied to the observed and model spectra under comparison (see below), so that the fit can be numerically evaluated on the same basis. The same argument also applies to our using the `pseudo' EWs as the fit quantities. 

Essentially the same procedures are adopted for the TARDIS model spectra, where the initial guesses for the edges of the absorption lines are taken from the corresponding `true' edges of the observed spectra. 
Applying the `observartionally' determined positions as the initial guess is justified, given our purpose of selecting model spectra that fit to the observed one; if a model spectrum has the characteristic wavelengths of the line profiles outside the searched range, such a spectrum is considered to provide a bad fit in the fitting procedure as described below. 
We then obtain the spectral properties for the model spectra, i.e., $\lambda_{\rm{minTAR}}(i)$, $\rm{EW_{TAR}}(i)$, $\rm{FWHM_{TAR}}(i)$, and $\rm{Col_{TAR}}$.

To evaluate the quality of the fit between an observed spectrum and a model spectrum, we then introduce weight parameters $W_{i}$ ($i=1-7$) for each element as shown below. We test various combinations of $W_{i}$ so that the following numerical fits reproduce the result of visual inspection reasonably well  (see below).
For (a), the quality of the fit for each line is defined as follows (which is smaller for a better fit);
\begin{eqnarray}
E_{a}(i) &=& \left( \frac{\lambda_{\rm{minTAR}}(i)-\lambda_{\rm{minobs}}(i)}{\Delta_{i}} \right)^{2} , \\
\Delta_{i} &=& \frac{\lambda_{\rm{minobs}}(i) \times 2,000_{\rm{km\ s^{-1}}}}{c}
\end{eqnarray}
where $c=3\times10^{5}$ km s$^{-1}$ (speed of light). $\Delta_{i}$ is a normalized constant; if the difference in the velocity of the absorption minimum between the observed spectrum and TARDIS model spectrum is 2,000 km s$^{-1}$, $E_{a}(i)$ is equal to 1.
By summing up $E_{a}(i)$ for all the lines in the fit ($i=1-7$) with the corresponding weight parameter, we obtain the following;
\begin{eqnarray}
E_{a}\__{\rm{tot}} = \sum_{i} E_{a}(i) \times W_{i} .
\end{eqnarray}

For (b), the quality of the fit is evaluated as follows;
\begin{equation}
E_{b}(i) = \left \{
\begin{array}{l} 
\left( \frac{\displaystyle {\rm nEW}_{\rm TAR}(i) - {\rm nEW}_{{\rm obs}}(i)}{\displaystyle {\rm nEW}_{\rm obs}(i)} \right)^2 \ (i \neq 6),\\
\\
\left( \frac{\displaystyle {\rm EW}_{\rm TAR}(i) - {\rm EW}_{{\rm obs}}(i)}{\displaystyle {\rm EW}_{\rm obs}(i)} \right)^2 \ (i=6),\\
\end{array}
\right.
\end{equation}
where
\begin{eqnarray}
    {\rm nEW}_{\rm k}(i) = {\rm EW}_{\rm k}(i) / {\rm EW}_{\rm k}(6). 
\end{eqnarray}
Namely, the EWs are normalized by the EW of \ion{Si}{2} 6355 $\rm{\AA}$ in the fit. Then,
\begin{eqnarray}
E_{b}\__{\rm tot} &=& \sum_{i} E_{b}(i) \times W_{i} .
\end{eqnarray}

For (c), the quality of the fit is evaluated as follows;
\begin{eqnarray}
E_{c}(i) &=& \left( \frac{{\rm FWHM}_{\rm TAR}(i) -  {\rm FWHM}_{\rm obs}(i)}{{\rm FWHM}_{\rm obs}(i)} \right)^2 ,\\
E_{c}\__{\rm tot} &=& \sum_{i}\  E_{c}(i) \times W_{i} .
\end{eqnarray}

For (d), the quality of the fit is evaluated as follows;
\begin{eqnarray}
Ed = \left( \frac{{\rm Col}_{\rm TAR} - {\rm Col}_{\rm obs}}{{\rm Col}_{\rm obs}} \right)^2
\end{eqnarray}

In adding all the contributions by items (a)-(d), we further introduce additional weight parameters, $W_{\alpha}^{'}$  ($\alpha$ = \{$a,b,c,d$\}). We then combine the fitting residuals computed for different features associated with different lines as follows;
\begin{eqnarray}
E_{\rm total} = \sum_{\alpha=a-d} W_{\alpha}^{'} \times E_{\alpha} .
\end{eqnarray}
We note that the values of $W_{\alpha}^{'}$ do not necessarily measure the relative importance of different items in the fit; they are coupled with the normalization in the corresponding $E_{\alpha}$. 

Through the above procedures, we find that a group of model spectra, which have too weak/shallow absorption lines but at the correct positions, are not always rejected, while they clearly do not provide a good match to the observed spectra by visual inspection. Rather than tuning the relative weight between the velocity minima (item a) and the EWs (item b) of individual line features, we find it easy to reject these models automatically by introducing another criterion of the `total EW' for each spectrum. We thus compute the sum of the EWs of individual line features in each model spectrum, and compare it to the corresponding value in the observed spectrum to fit. When the ratio of the model value to the observed one is smaller than the additional parameter $W^{'}_{e}$ ( taken to be 0.5), the model spectrum is judged to be `too smooth'. In this case, the $E_{\rm total}$ is added by 30; this value is arbitrary but set so that the model spectrum is essentially judged as unacceptable. 

With this additional constraint, the final value of $E_{\rm total}$ is given as follows;
\begin{eqnarray}
E_{\rm total} = \sum_{\alpha=a-d} W_{\alpha}^{'} \times E_{\alpha} + P,
\end{eqnarray}
\begin{equation}
P = \left \{
\begin{array}{l}
30 \ \ (\  \frac{\displaystyle \sum_{i} {\rm EW}_{\rm TAR}(i)}{\displaystyle \sum_{i} {\rm EW}_{\rm obs}(i)} < W^{'}_{e} \  ) \\
\\
0 \ \ (\  \frac{\displaystyle \sum_{i} {\rm EW}_{\rm TAR}(i)}{\displaystyle \sum_{i} {\rm EW}_{\rm obs}(i)} > W^{'}_{e} \ )
\end{array}
\right.
\end{equation}
We regard the quality of the fit better for a model with a smaller value of $E_{\rm tot}$.

The procedure here applies to given model grids and an observed spectrum, once the weight parameters are given. The weight parameters are introduced so that the numerical fit can reproduce the result of visual inspection, i.e., they must be set so that $E_{\rm total}$ is smaller for spectra that are judged to provide a better fit though visual inspection. This is difficult to numerically quantify, and necessarily involves subjectivity. We have done it through trial and error using the spectrum of SN 2011fe as test data. For a given model grid, we have selected a group of `good-fit' spectra through visual inspection. Then, we search for the combination of the weight parameters so that these models are selected in the top 3\% in the agreement with the observed spectrum. This procedure is repeated several time as the initial guess in the first step also uses information of the numerical fit. The values of the weight parameters thus obtained is shown in Table \ref{tab:weight}. We find that the same set of the weight parameters applies to the spectra of the other SNe investigated in the present sample, thus we use the same weight values throughout the analyses in the present work. This suggests that the procedure here is rather generic, and can potentially be expanded into a sample of SNe with different phases. 

\begin{deluxetable}{rc|rc}
\tablecaption{Weight parameters\label{tab:weight}}
\tablewidth{0pt}
\tablehead{
\colhead{index} & \colhead{$W_{i}$} & \colhead{index} & \colhead{$W_{\alpha}^{'}$}
}
\startdata
1	&	0.5	&	a	&	1	\\
2	&	0.5	&	b	&	0.06	\\
3	&	0.2	&	c	&	0.05	\\
4	&	0.3	&	d	&	0.01	\\
5	&	1	&	e	&	0.5	\\
6	&	1	&		&		\\
7	&	0.7	&		&		\\
\enddata
\tablecomments{These are set the same for all the objects.}
\end{deluxetable}

\begin{figure*}[t!]
\epsscale{1.05}
\plotone{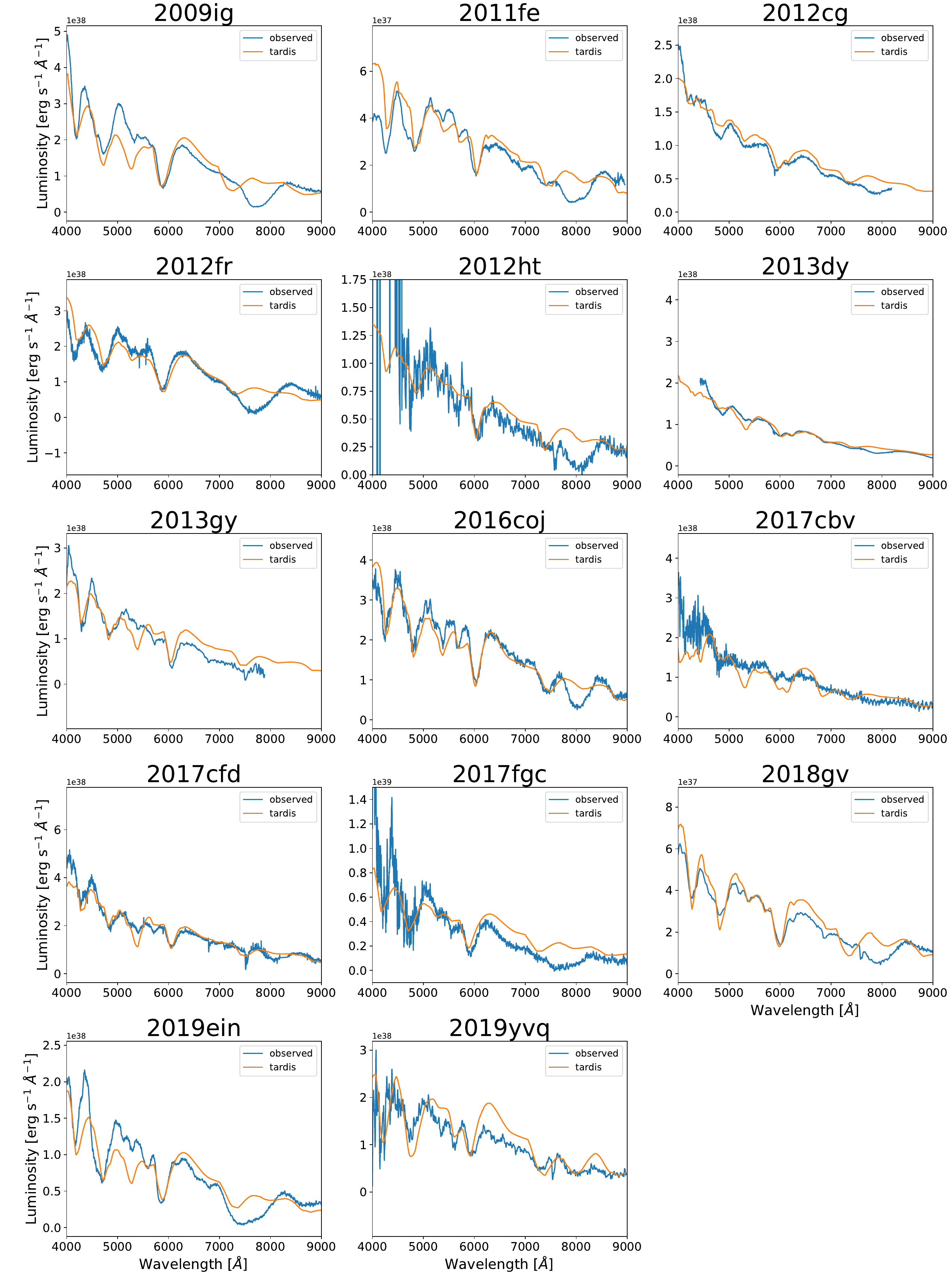}
\caption{Comparison between the observed spectrum and the TARDIS model spectrum for each object. In each panel, the blue line is for the observed spectrum, and the orange one is for the TARDIS model spectrum. \label{fig:all_plot}}
\end{figure*}

\section{Result}\label{sec:result}
Figure \ref{fig:all_plot} shows the results of the spectral synthesis calculations. The fitting procedure is constructed in a way such that a weight is given to the line features, i.e., the positions of the absorption minima, the depths and the widths, followed by the continuum color. This is partly because the normalization of the spectra is highly uncertainty due to the interstellar extinction, while the properties of the absorption lines are not sensitively affected. Furthermore, the line features contain a wealth of information in the spectral formation process, thus strongly reflecting the nature of supernovae. These best-fit spectra match the properties of  \ion{Si}{2} 6355 $\rm{\AA}$, the `w' feature of \ion{S}{2} 5606 $\&$ 5640 $\rm{\AA}$, as well as \ion{C}{2} 6578  (when it is robustly detected in the observed spectra). Note that in this study, we do not try to fit \ion{Ca}{2} near-infrared triplet because it is very sensitive to temperature and therefore to a detailed treatment of microphysics like non-LTE effect \citep{Kasen2006}. \ion{O}{1} 7774 \rm{\AA} is contaminated by \ion{Ca}{2}, and thus it is omitted from the fit either.

\begin{deluxetable*}{lcccccccc}
\tablecaption{The TARDIS parameters for the early-phase spectra of 14 SNe Ia\label{tab:tardis}}
\tablewidth{0pt}
\tablehead{
\colhead{Object} & \colhead{Time since} & \colhead{Luminosity} & \colhead{Photospheric} & \colhead{Outer} & \colhead{Photospheric} & \colhead{$\alpha$} &
\colhead{Fraction of C+O} & \colhead{Photospheric}\\ 
\colhead{} & \colhead{the explosion} & \colhead{} &
\colhead{velocity} & \colhead{velocity} &
\colhead{density*} & \colhead{} &
\colhead{unburnt layer} & \colhead{temperature} \\ 
\colhead{} &  \colhead{(day)} &
\colhead{(log($L_{\odot}$))} & \colhead{(km s$^{-1}$)} & \colhead{(km s$^{-1}$)} & \colhead{($10^{-16}$ g cm$^{-3}$)} & \colhead{} & \colhead{} & \colhead{(K)} 
}
\startdata
2009ig	&	3.9	&	8.55	&	19000(1271)	&	31000	&	$9.19^{12.31}_{5.22}$	&	4(1.6)	&	$0.00^{0.10}_{0.00}$	&	9933(2960)	\\
2011fe	&	2.5	&	7.80	&	14000(0)	&	24000	&	$9.94^{11.36}_{5.31}$	&	12(2.9)	&	$0.70^{0.10}_{0.20}$	&	9778(310)	\\
2012cg	&	2.9	&	8.35	&	19000(1285)	&	31000	&	$6.07^{15.73}_{4.38}$	&	10(2.1)	&	$0.70^{0.10}_{0.20}$	&	8124(0)		\\
2012fr	&	3.7	&	8.50	&	20000(771)	&	35000	&	$6.23^{75.47}_{5.75}$	&	8(2.4)	&	$0.00^{0.10}_{0.00}$	&	9873(6214)	\\
2012ht	&	3.4	&	8.15	&	15000(410)	&	21000	&	$3.82^{5.22}_{2.20}$	&	4(2.4)	&	$0.80^{0.00}_{0.30}$	&	9304(339)	\\
2013dy	&	2.6	&	8.45	&	16000(803)	&	30000	&	$3.45^{11.05}_{2.62}$	&	8(2.8)	&	$0.05^{0.05}_{0.00}$	&	12539(2471)	\\
2013gy	&	4.1	&	8.40	&	14000(542)	&	21000	&	$13.5^{6.2}_{4.2}$	&	8(3.2)	&	$0.05^{0.00}_{0.05}$	&	10206(246)	\\
2016coj	&	5.5	&	8.60	&	15000(183)	&	23000	&	$25.8^{18.5}_{10.7}$	&	10(2.4)	&	$0.00^{0.05}_{0.00}$	&	9447(233)	\\
2017cbv	&	3.0	&	8.45	&	17000(1135)	&	31000	&	$13.6^{15.1}_{7.2}$	&	8(1.5)	&	$0.50^{0.20}_{0.00}$	&	11558(1560)	\\
2017cfd	&	4.9	&	8.70	&	14000(594)	&	21000	&	$37.1^{22.5}_{14.2}$	&	10(2.2)	&	$0.01^{0.09}_{0.00}$	&	11260(690)	\\
2017fgc	&	6.1	&	8.90	&	21000(702)	&	30000	&	$17.4^{5.9}_{4.3}$	&	6(1.5)	&	$0.00^{0.30}_{0.00}$	&	9049(232)	\\
2018gv	&	2.5	&	7.80	&	14000(756)	&	26000	&	$4.92^{4.16}_{2.26}$	&	6(1.8)	&	$0.70^{0.20}_{0.20}$	&	9445(429)	\\
2019ein	&	2.6	&	8.25	&	20000(1111)	&	30000	&	$5.54^{9.06}_{3.44}$	&	8(2.0)	&	$0.00^{0.05}_{0.00}$	&	9977(2786)	\\
2019yvq	&	4.8	&	8.40	&	19000(780)	&	28000	&	$27.5^{18.5}_{11.0}$	&	10(2.1)	&	$0.00^{0.30}_{0.00}$	&	8799(513)	\\
\enddata
\tablecomments{ The numbers in parentheses correspond to 1-$\sigma$ statistical uncertainties. The subscript and superscript values represent the lower and upper 1-$\sigma$ statistical error, respectively. Photospheric density is normalized to the corresponding value on 20 days since the explosion.}
\end{deluxetable*}

\begin{figure*}[t!]
\epsscale{1.1}
\plotone{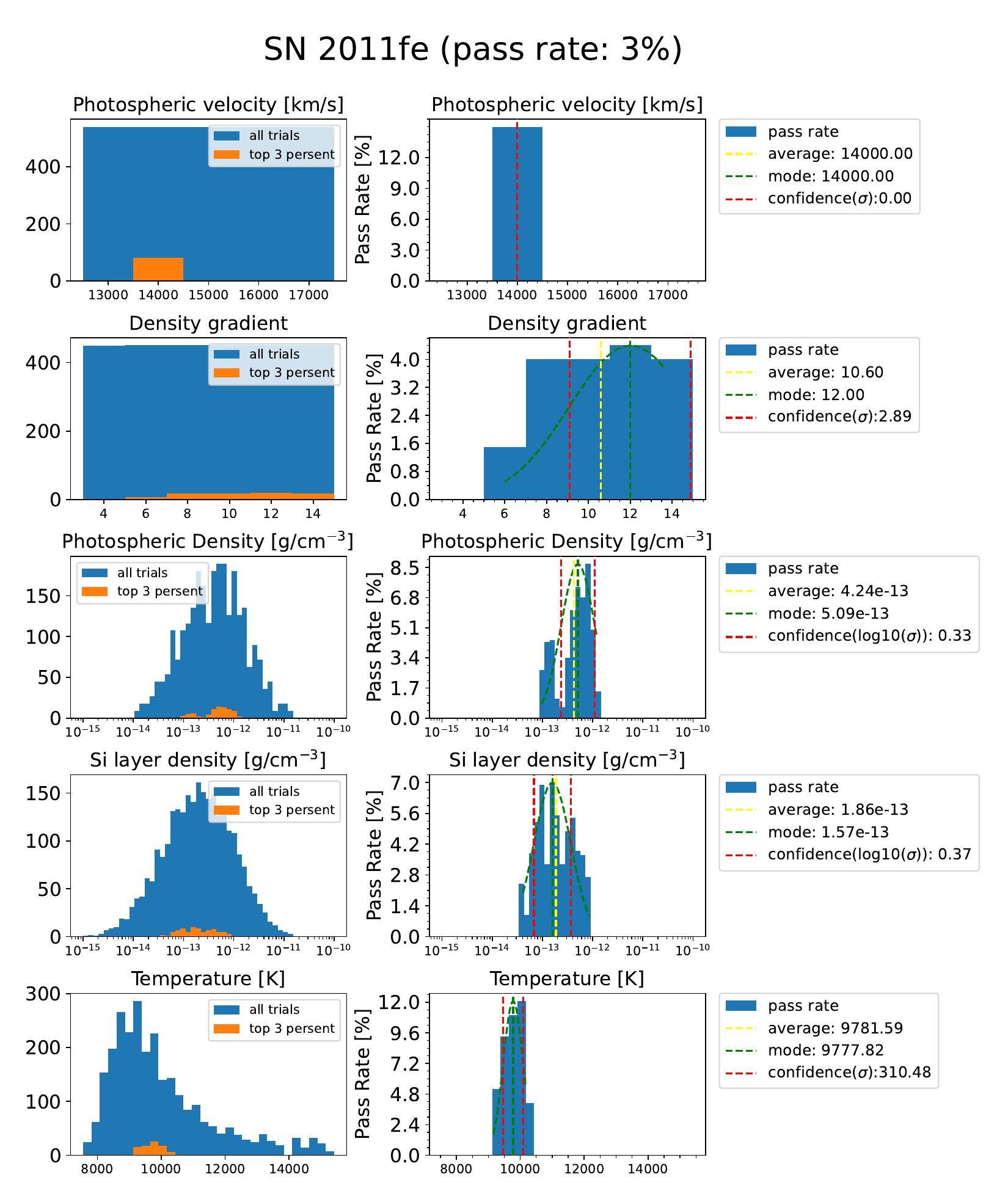}
\caption{The left panels show the distributions of the number of all the trials (blue) and the number of the acceptable models (those in the top  3 percent in the agreement with the observed one; orange), for each parameter. 
 Right panels show the distribution of the `pass rate', with the 1$\sigma$ confidence interval (red dashed line).}
\label{fig:2011fe_tardis}
\end{figure*}

Table \ref{tab:tardis} shows the parameters for the best-fit model selected for each object (Fig. \ref{fig:all_plot}).
These models are selected as follows. For each object, we rank all the models according to the value of $E_{\rm total}$ in the fit which is numerically (and automatically) obtained. The models that are in the top 3\% in the fitting rank (the smaller $E_{\rm total}$) are grouped as the best-fit `candidates' (or, the acceptable models). 
Then we calculate the `pass rate’ for each parameter, which is defined as the number of the acceptable models (i.e., the best-fit candidates within the top 3$\%$) divided by the number of all the trials for each parameter value under consideration. Since the pass rate corresponds to a probability density, the `best-fit' value for that parameter is considered to be near the peak
of the pass-rate distribution. We decide to adopt the `mode' in the distribution for each parameter as the corresponding best-fit value. 

Since the numerical evaluation of the fit is aimed to be a simplified (and automatic) algorithm to `mimic' the visual inspection, it sometimes cannot recognize apparently slight absorption lines such as \ion{C}{2}. 
We thus perform an additional selection procedure based on visual inspection, as follows. The best-fit values for the parameters `except for the fraction of the CO layer' are taken as the mode values in the pass-rate distribution for each parameter (see above). For this parameter set, we pick up a model series for which only the fraction of the CO layer is varied. We then select the best-fit model by visual inspection between this model series and the observed spectrum under consideration. This way, we constrain the fraction of the CO layer separately from the other parameters. 

 Note that the best-fit models shown in Figure \ref{fig:all_plot} are not always the ones with the smallest $E_{\rm total}$, following the steps of the selection procedures as described above. We emphasize that our procedure provides a way to measure an error and uncertainty associated with each parameter by inspecting the range of the parameter sets in the acceptable or unacceptable groups.  Our procedure thus enhances reproducibility; the evaluation of the fit is not too arbitrary nor too biased by the experience of individual researchers, as the best-fit value for each parameter (except for the fraction of the CO layer) is determined by the automatic evaluation processes (for which the details of the algorithm are presented and can be reproduced); the fraction of the CO layer involves visual inspection, but it is done with the other parameters automatically fixed and thus the degree of freedom in the fit has been already reduced substantially.

\begin{figure*}[t!]
\epsscale{1.29}
\plotone{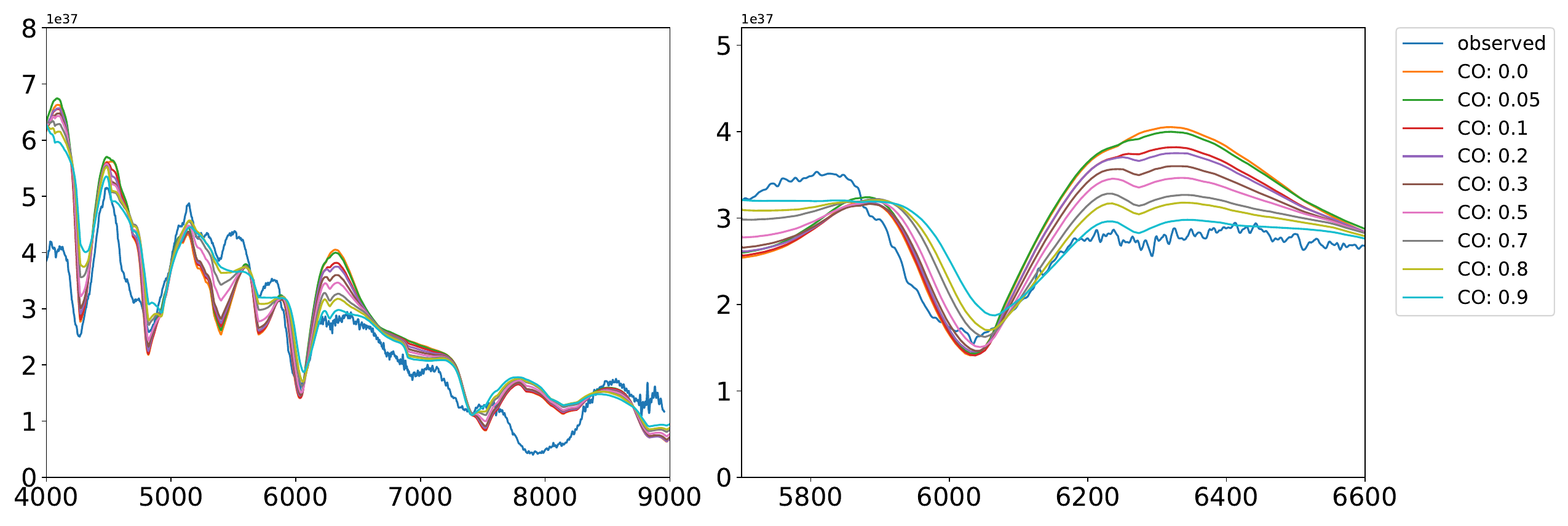}
\caption{ The left panel shows the spectra with the `best-fit' value for each parameter except for the fraction of CO layer, showing how the amount of the CO layer affects the spectrum. The right panel is an expanded view around the \ion{C}{2}.}
\label{fig:2011fe_COspec}
\end{figure*}

\subsection{An Example of the best fit model: SN 2011fe}\label{sec:2011fe}
In this work, since we use the numerical test for the quality of fit, we can discuss uniqueness and uncertainty of the best-fit parameters. 
To illustrate this point, in this section we introduce the results of our fitting algorithm for a specific example, SN 2011fe.  In the second step for SN 2011fe, we calculate 2699 models in total, and define the group of the models that belong to the top 3 percent in the fitting agreement that pass the automatic screening (i.e., `the best-fit candidates' or `acceptable models'). 
The result is shown in Figure \ref{fig:2011fe_tardis}. 
In the left panels, we show the distributions of the number of models we have calculated (blue) and the number of the models that pass the screening (orange), for some selected parameters. 
In the right panels, we show the distribution of the `pass rate', defined as the number of the acceptable models divided by the number of all the trials for each parameter value.  
We also calculate an error for each parameter, which is defined as 1$\sigma$ confidence interval from the mode in the pass-rate distribution.

With the best-fit parameters (except for the fraction of the CO layer) fixed as the mode values in the pass-rate distributions, we then perform the additional selection regarding the \ion{C}{2} absorption line based on visual inspection. Figure \ref{fig:2011fe_COspec} shows spectra for this final model sequence (with varying the fraction of the CO layer) as compared to the spectrum of SN 2011fe. Through visual inspection, we regard the best-fit model as the one with the fraction of the CO layer being 0.8. As mentioned above, the best-fit model is not necessarily the one with the smallest E$_{\rm total}$. The best-fit model adopted for SN 2011fe through the above processes is 
ranked as 19th in terms of $E_{\rm total}$ among the 2699 models (i.e., in the top 0.7\%).

According to Figure \ref{fig:2011fe_tardis}, we find the photospheric density, the location of the photosphere ($v_{\rm ph}$), and the photospheric temperature are very well determined. In general, the density and the abundance can degenerate. 
However, by focusing on the very early-phase spectra, we can place a reasonable assumption that the abundance is well represented by the C+O unburnt layer and/or carbon burning layer. This is how the photospheric density and the abundance are separately constrained well in the present study.
On the other hand, it turns out that the density gradient is not so strongly constrained. This is because information near the photosphere mainly contributes to the shape the spectral feature. As the photospheric temperature is well constrained by the ionization and excitation conditions, together with the continuum color, the luminosity is tightly constrained by the overall flux level.

\subsection{A summary of the fits to all the sample}\label{sec:model_summary}
Table \ref{tab:numerical_analysis} summarizes the results of the numerical evaluation for all the objects. NV SNe tend to have small values for minimum $E_{total}$ except for SN 2012fr, which we discuss further in Section \ref{sec:2012fr}. 
On the other hand, HV SNe and 1999aa-like SNe tend to show a relatively high value in the minimum $E_{\rm total}$, as compared to NV SNe. This might suggest that some of the assumptions in the model construction are worse in the high-velocity ejecta found in these subclasses than in the low-velocity ejecta found in NV SNe. One possibility is the assumption of the single power-law density distribution, which we plan to investigate in the future. 

\begin{deluxetable}{rccccc}
\tablecaption{Numerical analysis\label{tab:numerical_analysis}}
\tablewidth{0pt}
\tablehead{
\colhead{object} & \colhead{The number of} &\colhead{The number of} & \colhead{minimum}
\\
\colhead{} & \colhead{all the trials} & \colhead{the top 5\% models} & \colhead{$E_{total}$} \
}
\startdata
2009ig	&	2915	&	87	&	2.6777 	\\
2011fe	&	2699	&	80	&	0.2182 	\\
2012cg	&	2577	&	77	&	1.5411 	\\
2012fr	&	1889	&	56	&	3.2501 	\\
2012ht	&	1889	&	56	&	0.8045 	\\
2013dy	&	1889	&	56	&	0.4637 	\\
2013gy	&	1919	&	57	&	1.0206 	\\
2016coj	&	1919	&	57	&	0.5107 	\\
2017cbv	&	1943	&	58	&	2.2982 	\\
2017cfd	&	2159	&	64	&	0.2812 	\\
2017fgc	&	2159	&	64	&	3.2283 	\\
2018gv	&	2302	&	69	&	0.6455 	\\
2019ein	&	1917	&	57	&	1.9852 	\\
2019yvq	&	2303	&	69	&	1.1558 	\\
\enddata
\end{deluxetable}

In any case, given that the the model parameters (e.g., the density, velocity, and the composition) are reasonably well constrained under the present model framework (Section \ref{sec:2011fe}), it is unlikely that these are affected much by introducing a more complicated model structure. In addition, the difference in the distribution in the minimal $E_{total}$ is just indicative, and further investigation of this possible difference will require enlarging the observational sample. 

\subsection{Density structure\label{sec:density}}

\begin{figure*}[t!]
\centering
\epsscale{1.2}
\plotone{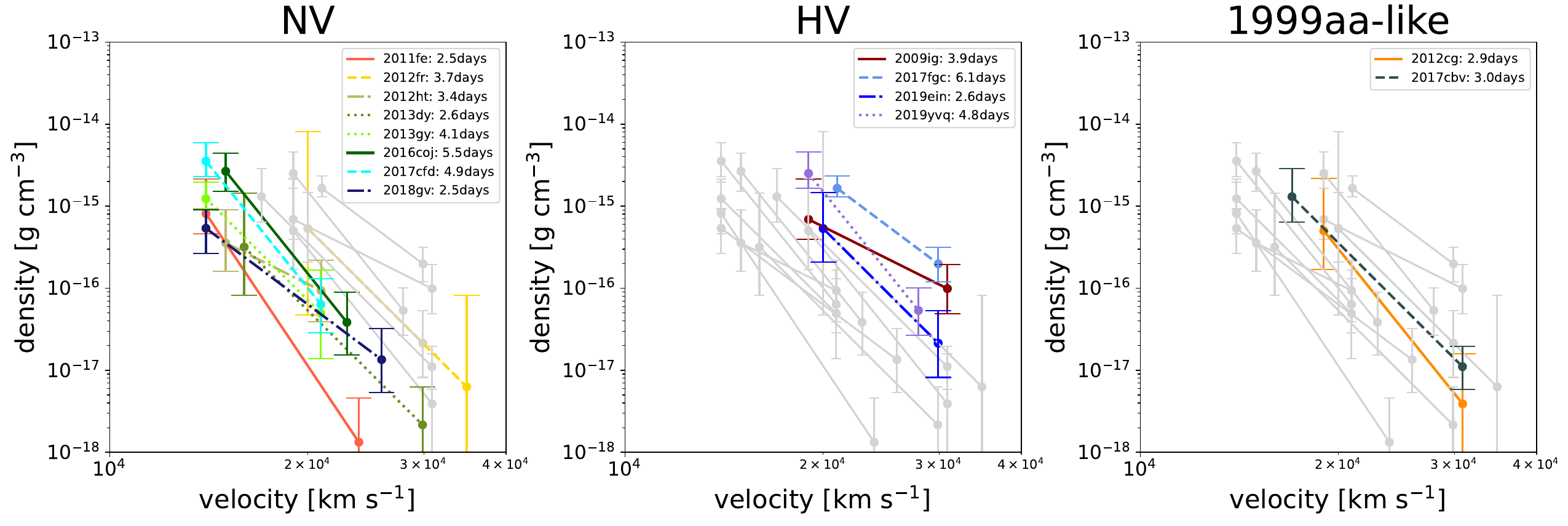}
\caption{Density structure of each object based on the best-fit TARDIS model. The error bars in the density attached at the innermost and outermost velocities for each object  represent the 1$\sigma$ oonfidence interval (see the main text).}
\label{fig:density}
\end{figure*}

Figure \ref{fig:density} shows the density structures of the best-fit models for the 14 SNe Ia with the estimated error/uncertainty, as obtained by the TARDIS modeling. The figure is separated to three panels for different subclasses.  
In the NV SNe, the photospheric velocity is all confined at $\sim$15,000 km\ s$^{-1}$. The outermost velocity, which corresponds to the region to which the carbon burning extends, shows variation from $\sim$20,000 km s$^{-1}$ to $\sim$25,000 km s$^{-1}$. The ejecta density also shows diversity nearly by an order of magnitude. 

The velocity range found for the HV SNe and 1999aa-like SNe is largely overlapping; 
the photospheric velocity and the outermost velocity are $\sim$20,000 km s$^{-1}$ and $\sim$30,000 km s$^{-1}$, respectively, which are both higher than those of the NV SNe. The high velocity derived for the HV SNe is also consistent with the high velocity seen in the Ca II NIR triplet \citep{li2021}.
The ejecta densities span the similar range with that covered by the NV SNe, between $\sim 10^{-15}$ (at the photospehre) and  $10^{-17}$ g cm$^{-3}$ (at the outermost edge), once scaled at 20 days since the explosion. 
As a combination of the HV SNe and 1999aa-like SNe, the range of the density `scale', i.e., the density at the photosphere scaled at 20 days, is also similar to  that found for the NV SNe (about an order of magnitude). 
Interestingly, we can discern the potential difference in the densities between the HV SNe and 1999aa-like SNe; the typical density of the HV SNe is higher than that of the 1999aa-like SNe. 
 This is discussed further in Section \ref{sec:population}.
In summary, we find that the outermost density structures are distinct for different sub-classes of SNe Ia.

\subsection{Composition structure\label{sec:composition}}
\begin{figure}[t!]
\epsscale{1.2}
\plotone{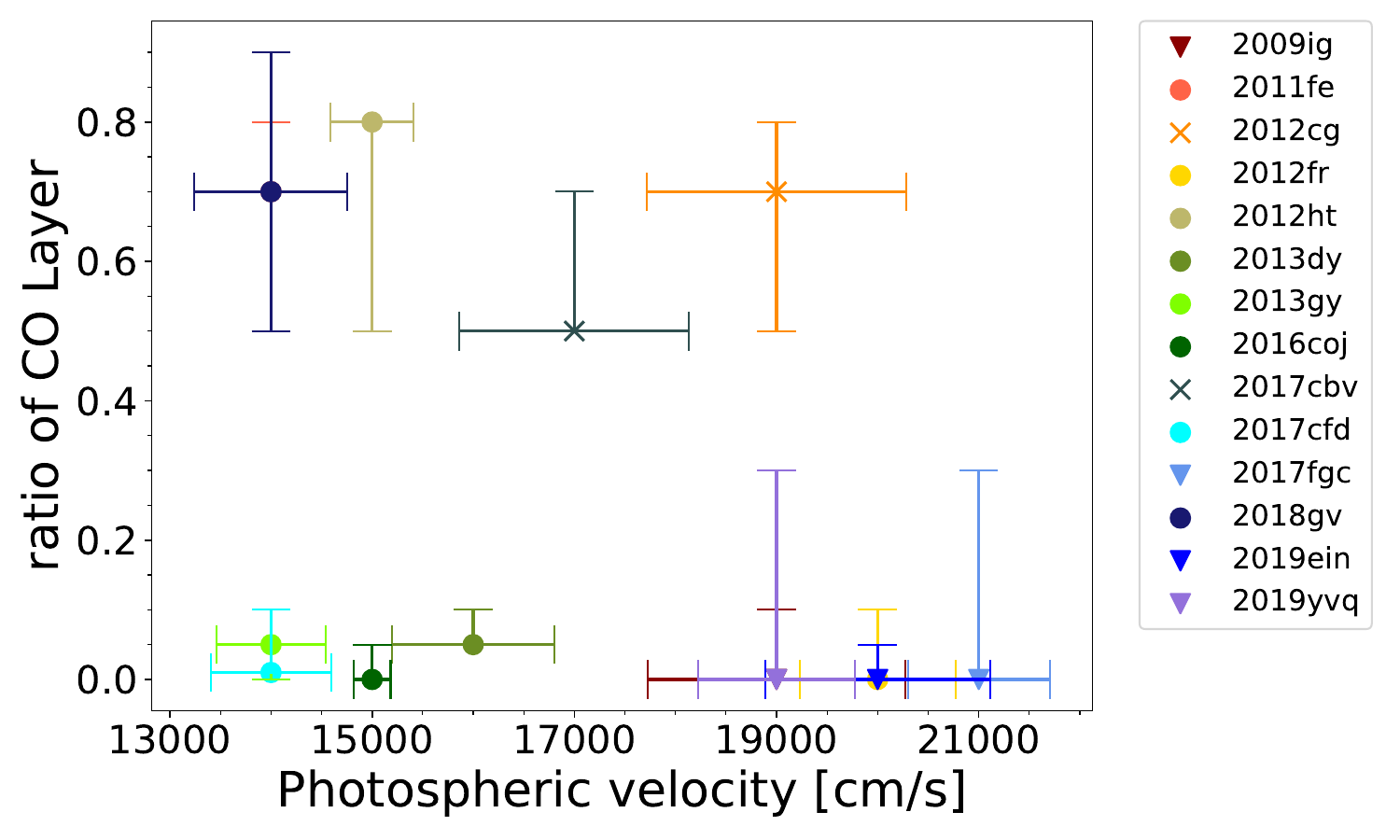}
\caption{The fraction of the unburnt C+O layer to the total composition. Circles represent the NV type, inverse triangles are for the HV type, and crosses correspond to the 1999aa-like type.
The error bars are obtained by the additional visual-inspection process to constrain the fraction of the CO layer, where only this parameter is varied while the other parameters are fixed to the best-fit values (see Section \ref{sec:2011fe}).
}
\label{fig:carbon_fraction}
\end{figure}

Figure \ref{fig:carbon_fraction} shows the fraction of the unburnt C+O layer contained in the outermost ejecta as constrained by the earliest spectra available for each object. 
In the 1999aa-like objects, we find that the fraction of the unburnt C+O layer is $\sim$ 0.6. This means that the outermost layer is basically the unburnt C+O layer, and it is mixed by a little amount of the carbon burning layer. The situation is the opposite for the HV SNe, for which the fraction of the C+O layer is essentially zero. Namely, the outermost layer of the HV SNe is represented by the carbon burning composition \citep[see also][]{li2021}.
We find that the NV SNe are divided into two groups in the composition of the outermost layer; one with a large fraction of the unburnt C+O layer and the other dominated by the carbon burning layer\footnote{ Note that we select the model showing the C II for SN 2012ht, although the large noise level seen in the observed spectrum does not allow clear identification of the C II; we see the C II in this spectrum once it is smoothed, and the line is clearly seen in a spectrum taken 2 days after.}.  The relation between the different sub-classes and the fraction of the CO layer is further discussed in Section \ref{sec:population}.

Table \ref{tab:carbon_mass} shows the carbon masses contained in the best-fit models. Note that the values here are the lower limits because there could be additional carbon either above or below the velocity range constrained by the present models.
First, there could be a pure C+O unburnt layer outside the ejecta shell modeled in this study. However, this contribution is probably negligible; we model the very early-phase spectra, and the amount of material above the outermost velocity in our models should be very small given the steep density structure. On the other hand, potential existence of carbon below the photosphere is not rejected by this study. This contribution should be negligible for the objects showing the carbon-poor composition in the outermost layer. For the objects showing a large fraction of the unburnt carbon layer (1999aa-like and some NV SNe), the carbon masses in Table \ref{tab:carbon_mass} are likely underestimated.

Those showing the carbon-rich composition (1999aa-like SNe and some NV SNe) have the carbon masses of $\gsim 0.001 - 0.01 M_{\odot}$ (noting that these are the lower limits). On the other hand, those showing the carbon-poor composition (HV SNe and other NV SNe) have the carbon masses of $\lsim 0.001 M_{\odot}$ (with only exception of SN 2019ein). Therefore, it is seen that these two groups, divided by the content of the unburnt carbon layer, are also distinct in the carbon masses. 

\begin{deluxetable}{lcc}
\tablecaption{Carbon mass\label{tab:carbon_mass}}
\tablewidth{0pt}
\tablehead{
\colhead{object} & \colhead{Carbon mass} & \colhead{Total mass}
\\
\colhead{} & \colhead{($M_{\odot}$)} & \colhead{($M_{\odot}$)}
}\
\startdata
2009ig	&	0		&	0.059693535	\\
2011fe	&	0.002747892	&	0.008292515	\\
2012cg	&	0.00524938	&	0.015841442	\\
2012fr	&	0		&	0.02665105	\\
2012ht	&	0.004208668	&	0.011113205	\\
2013dy	&	0.000197021	&	0.008323906	\\
2013gy	&	0.000458434	&	0.019368311	\\
2016coj	&	0		&	0.040493987	\\
2017cbv	&	0.009617398	&	0.040632383	\\
2017cfd	&	0.000204989	&	0.043302796	\\
2017fgc	&	0		&	0.109386767	\\
2018gv	&	0.00454342	&	0.013711013	\\
2019ein	&	0		&	0.02441424	\\
2019yvq	&	0		&	0.075874934	\\
\enddata
\tablecomments{The carbon mass and the total mass in the outermost ejecta studied in this work.}
\end{deluxetable}

\section{Insights into the nature of SNe Ia}\label{sec:implication}

\begin{figure}[t!]
\centering
\epsscale{1.2}
\plotone{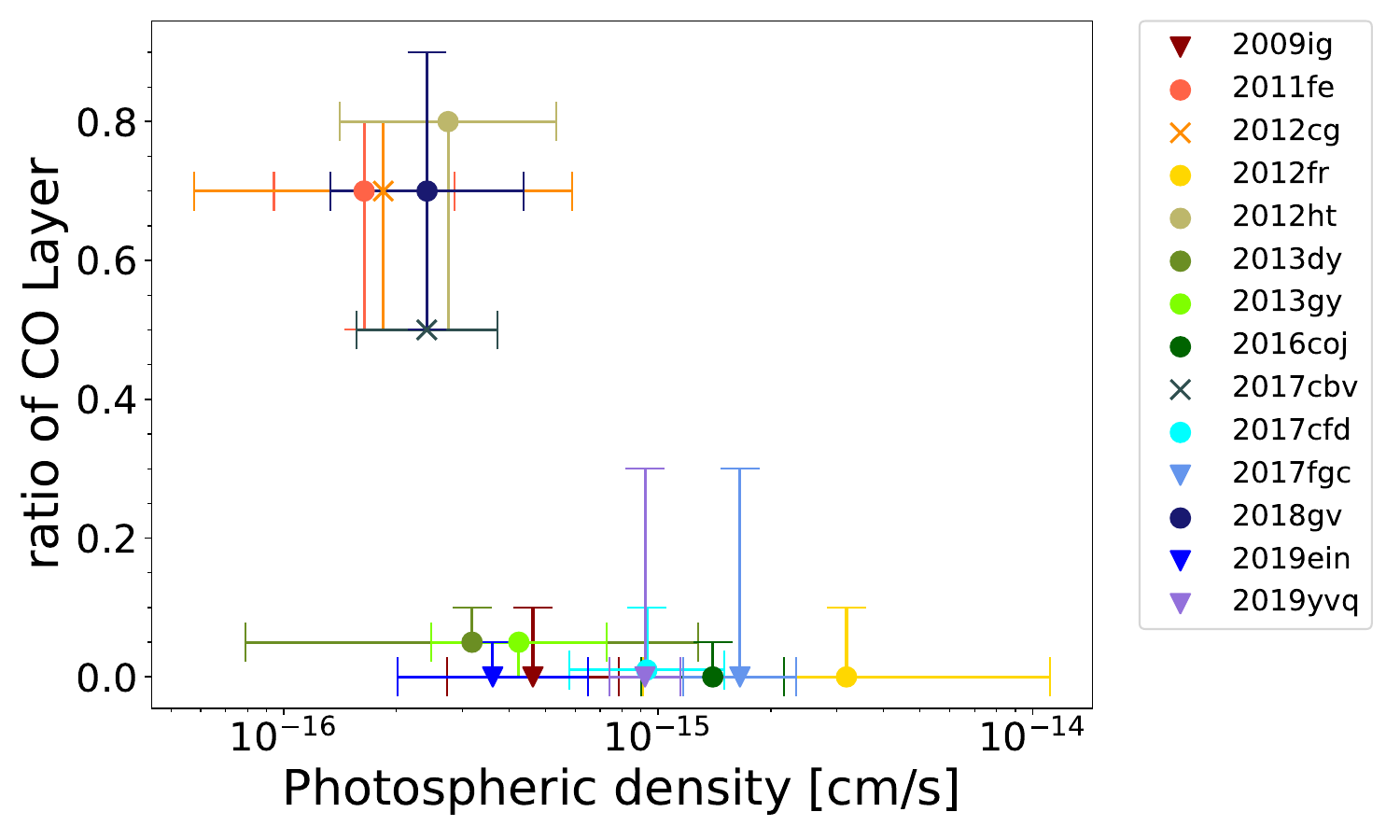}
\caption{ Characteristic density as compared to the fraction of the CO layer for the 14 SNe. The density is scaled at $16000$ km s$^{-1}$ (for NV SNe) or $21000$ km s$^{-1}$ (for HV SNe and 99aa-like SNe). 
}
\label{fig:density_vs_co}
\end{figure}

\begin{figure*}[t!]
\centering
\epsscale{0.57}
\plotone{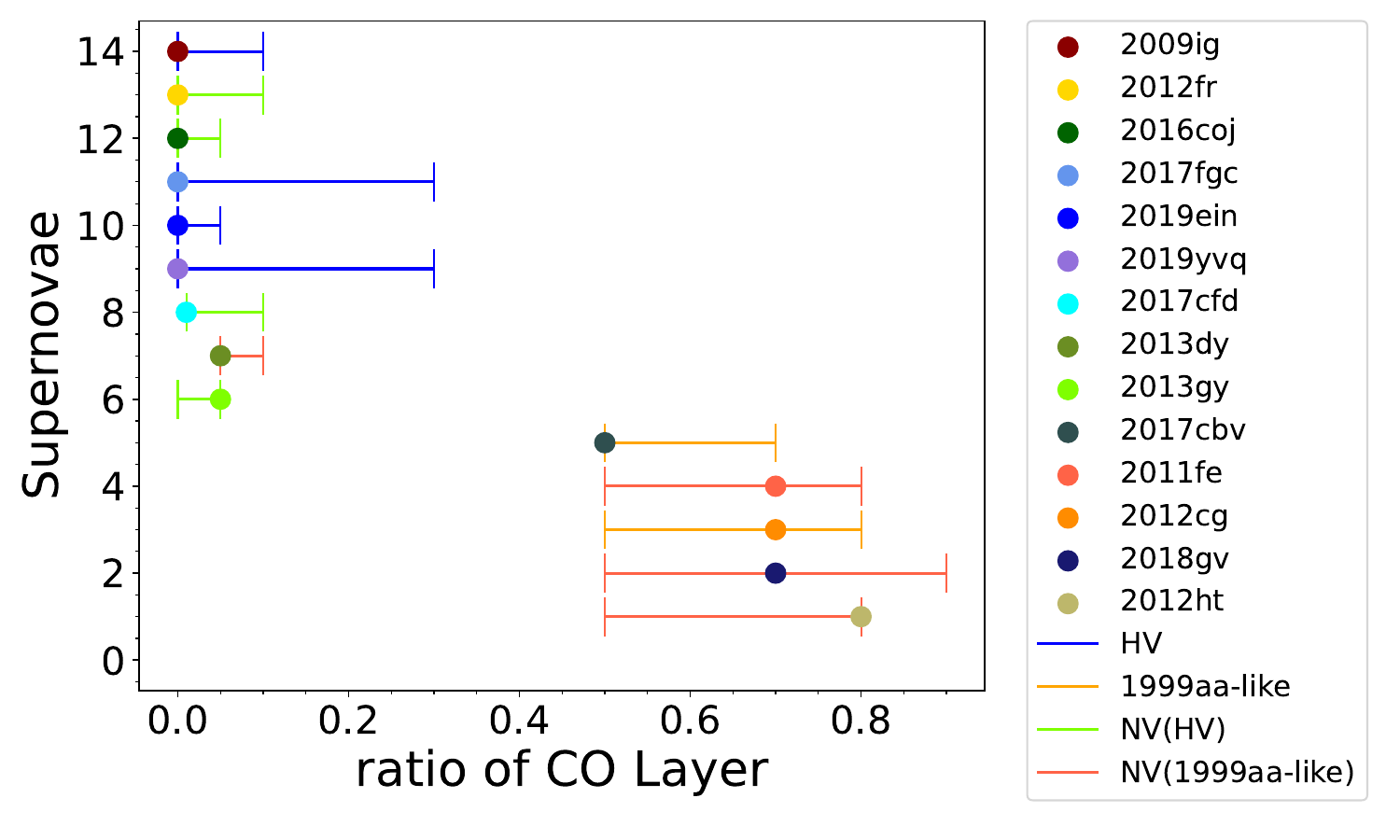}
\plotone{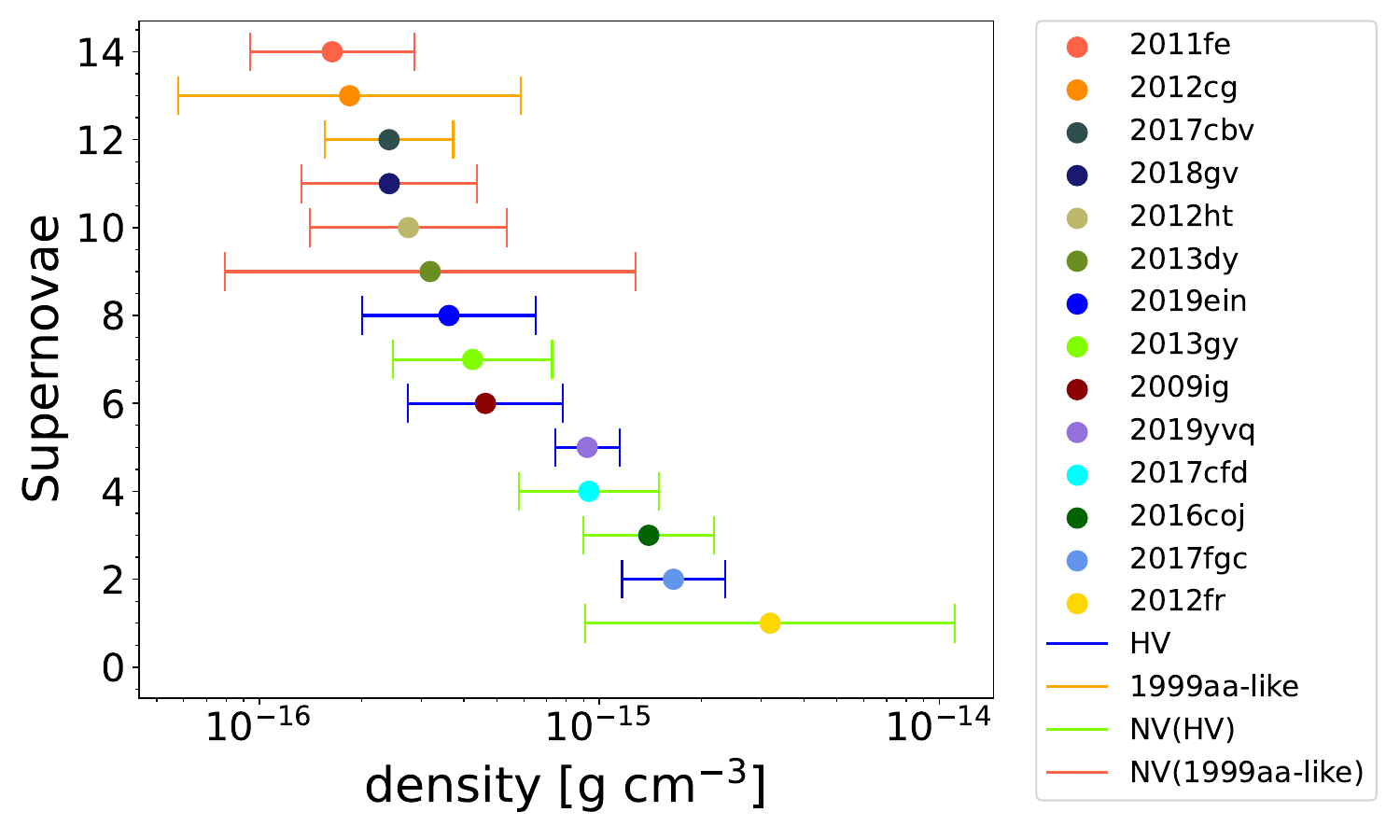}
\caption{ The fraction of the CO layer (left) and the characteristic density (right) as shown in the increasing order from top to bottom. The different sub-classes are marked by different colors for the error bars; 99aa-like SNe (orange) and HV SNe (blue), and `C-rich' NV SNe (red) and `C-poor' NV SNe (green). 
}
\label{fig:two_groups}
\end{figure*}

\subsection{Two sequences in SN Ia populations?}\label{sec:population}
 In Sections \ref{sec:density} and \ref{sec:composition}, we have shown that HV SNe and 99aa-like SNe have different characteristics in the outermost density and composition, despite the similar velocity range found for the outermost ejecta. 
In Section \ref{sec:composition}, we have further shown that NV SNe could be divided into two classes based on the composition structure in the outermost ejecta; one dominated by the unburnt C+O composition and the other showing little trace of the unburnt C+O layer. The division between the two classes is also related to the outermost density (Section \ref{sec:density}).

To further quantify these points, Figure \ref{fig:density_vs_co} shows the characteristic density (scaled for the same epoch for all the SNe), as compared to the fraction of the CO layer for the 14 SNe. The characteristic density is defined as those at $16000$ km s$^{-1}$ (for NV SNe) or $21000$ km s$^{-1}$ (for HV SNe and 99aa-like SNe), i.e., just above the photosphere. Figure \ref{fig:two_groups} summarizes how the fraction of the CO layer and the characteristic density are related to different sub-groups; HV SNe vs. 99aa-like SNe, and further NV SNe are divided into two sub-classes (C-rich and C-poor; see below). According to the Student's t-test, the chance probability with which the distribution of the CO fraction of the HV SNe and that of the 99aa-like SNe is derived from the same parent population is $\sim 6.0$\%. The same chance probability for the distribution of the characteristic density is also $\sim 6.0$\%. While the sample size is still small, the difference between HV SNe and 99aa-like SNe is clearly indicated. 

The density of the NV SNe with little unburnt C+O composition is as high as that found for HV SNe. This group may be regarded as a low-velocity analog of HV SNe, which also show both high density and little contamination of the unburnt C+O layer. 
On the other hand, the density seen in the group of the NV SNe having a large amount of the unburnt C+O in the outermost layer is not as high as that of HV SNe. Interestingly, these features exhibited by this carbon-rich NV group are similar to those of 1999aa-like SNe, except for the difference in the velocity; this group may be regarded as a low-velocity analog of 1999aa-like SNe. 
 By dividing the NV SNe into the two classes based on the fraction of the CO layer (those with the fraction being $> 0.5$ and the others with $< 0.1$\footnote{ We however group SN 2013dy into the `carbon-rich' 99aa-like analog despite its low CO fraction, following the discussion in Section \ref{sec:2013dy}.}), we see the clear difference between the two groups (Figure. \ref{fig:two_groups}); In terms of the fraction of the CO layer, the possibility that the two groups are derived from the same parent population is $\sim 2.5$\%, and the same probability is $\sim 9.3$\% for the characteristic density.

Therefore, we propose that NV SNe may be divided into two populations, and that SNe Ia might be divided into two sequences; one is the NV--HV sequence and the other is the NV--1991T/1999aa-like sequence. The former sequence is characterized by the high density and little amount of the unburnt C+O later in the outermost ejecta (which might be linked to a similar suggestion by \citealt{li2021} based on observational properties); the latter sequence has the lower density than the NV--HV sequence, and the outermost layer is dominated by the unburnt C+O composition. 
Further investigating other observational properties in view of the proposed two populations should be interesting; for example, the HV SNe show a high polarization level around the maximum light \citep[e.g.,][]{Maeda2010}, and indeed SNe Ia might be divided into two groups in which one shows a high polarization degree (including HV SNe) and the other shows a low polarization degree (including 1991T-like SNe) \citep{meng2017}.

\subsection{SN 2012fr as a transitional case in the NV--HV sequence ?}\label{sec:2012fr}
We find two outliers in the NV SN category, for which the outermost velocity is substantially higher than the other NV SNe Ia, noting that the NV classification is based on the maximum-light spectra. 
SN 2012fr is one such example. Despite its being classified as a NV SN, \citet{Childress2013} showed that \ion{Si}{2} 6355 and \ion{Ca}{2} infrared triplet seen in SN 2012fr consist of a strong high velocity feature (HVF) component in the early phase, which disappears toward the maximum phase. The maximum-light spectra are dominated by a slower, `photospheric' component and resemble those of other NV SNe. The density and composition structures derived in this study are very close to those of the HV SNe. The density in the outermost layer of SN 2012fr is as high as that derived for HV SNe, and the composition has little trace of the unburnt C+O layer as is similar to the case of HV SNe.

These properties suggest that the structure of the outermost layer of SN 2012fr, which creates emissions in the very early phase, is similar to that of the HV type. As one moves inward, the inner structure may become close to that of the NV type. As such, we suggest that SN 2012fr is placed as a transitional object between the NV and HV classes in the NV--HV sequence.

\subsection{SN 2013dy as a transitional case in the NV--1991T/1999aa-like sequence?}\label{sec:2013dy}
The other outlier in the NV class is SN 2013dy.
While SN 2013dy is classified as the NV SN based on the maximum-phase spectra, the outermost density and composition structures derived in this study are very close to those of the 1991T/1999aa-like type. Unlike the other NV SNe (except for SN 2012fr), the ejecta extend to $\sim$30,000 km s$^{-1}$, which is comparable to 1991T/1999aa-like SNe. The density is overall low; it is at the lowest side among NV SNe at the photosphere, and it is comparable to 1991T/1999aa-like SNe in the outermost region. The fraction of the unburnt C+O layer in the outermost layer is $\sim$ 0.05, which is not as high as those of 1999T-like but  we note that those derived for HV SNe are essentially zero.

\begin{figure*}[t!]
\epsscale{0.62}
\hspace{-1cm}
\plotone{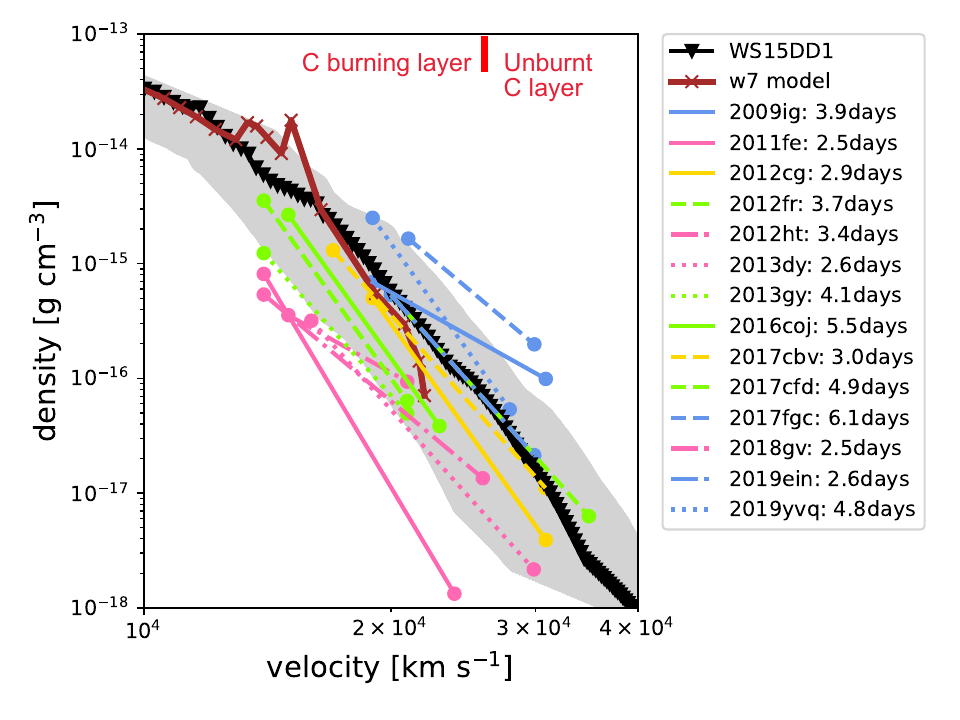}
\hspace{-0.5cm}
\plotone{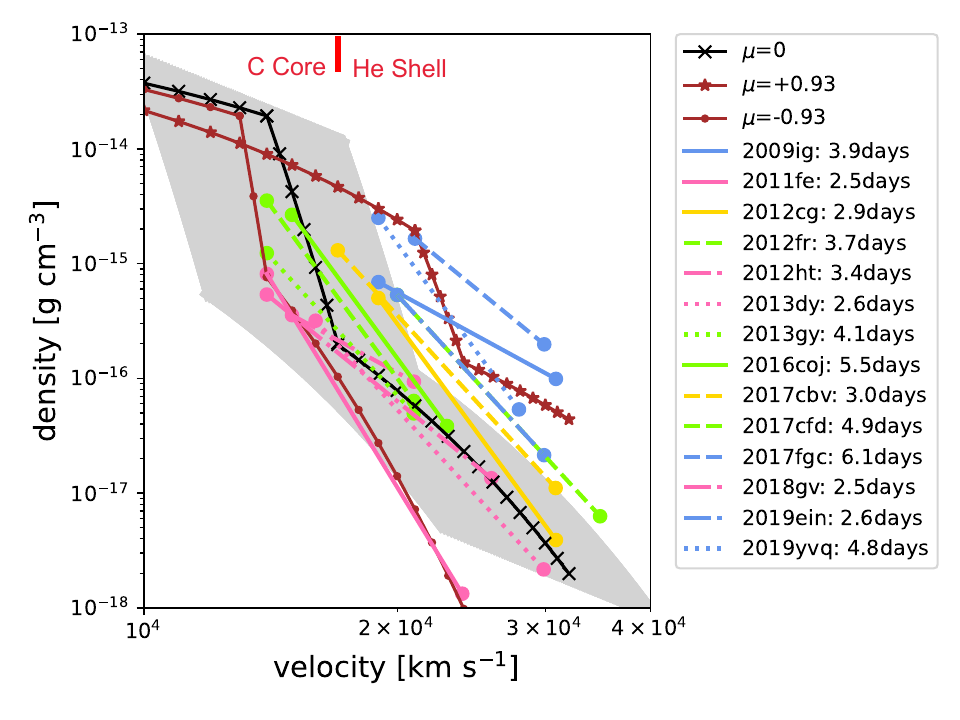}
\caption{ Comparison of the density structures obtained by TARDIS to some theoretical models. The left panel shows W7 model \citep[brown;][]{nomoto1984}, WS15DD1 model \citep[black;][]{Iwamoto1999}; the right panel shows the double detonation model of \citet{Shen2021}, for three choices of the viewing directions (black for $\mu= 0$ and brown for $\mu = \pm 0.93$). 
Gray-shaded region shows the trajectory of WS15DD1 model (left) or the double detonation model for an average viewing direction ($\mu = 0$, right), when the kinetic energy is changed by $ \pm 50 \%$. Vertical red line shows the boundary between different layers (see labels). Different colors are used to show the ejecta structures for individual SNe, depending on the subclass they belong to; blue lines for the HV SNe, yellow lines for the 1999aa-like SNe. The NV SNe with a large and little amount of the unburnt carbon layer are denoted by pink and green lines, respectively.
}
\label{fig:density_model}
\end{figure*}

These properties suggest that the structure of the outermost layer, as traced in the very early phase, may be similar to those of the 1991T-like type or may represent an intermediate case between 1991T-like and NV SNe. As one moves inward, the inner structure may become close to that of the NV type. We note that the photospheric velocity of SN 2013dy in the earliest phase is typical of NV SNe, which is lower than 1991T/1999aa-like SNe. This is consistent with the density structure of SN 2013dy; while the density is comparable to those of 1991T/1999aa-like SNe in the outermost layer, it is overall lower than those of 1991T/1999aa-like SNe toward lower velocities. It is therefore expected that the photosphere is not kept at a high velocity in SN 2013dy.

\subsection{Possible relations to the explosion mechanisms}
The two possible sequences, the 1991T/1999aa--NV and the HV--NV sequences, indicate that there might exist two distinct progenitor  channels and/or explosion scenarios for SNe Ia. In this section, we first discuss the pros and cons of the two popular models, the delayed-detonation model (and the closely related W7 model) and the double-detonation model. The possible relation of each scenario to the different SN Ia populations is discussed.  While the content of this section is speculative, we believe that it provides a useful guide for further development especially in the modeling activity.

Figure \ref{fig:density_model} shows the comparison between the density structure obtained by the TARDIS spectral synthesis and the $M_{\rm ch}$ WD models for which the thermonuclear runaway is triggered in the central region of the WD; a pure-deflagration model W7 \citep{nomoto1984} and a delayed detonation model WS15DD1 \citep{Iwamoto1999}. These two models have similar overall properties in the WD mass (near $M_{\rm Ch}$), the $^{56}$Ni mass ($\sim 0.6 M_\odot$) and the explosion energy ($\sim1.5\times10^{51}$ erg). 

For the density structure, the W7 model would not explain the high velocity seen in HV SNe and 1999aa-like SNe. The delayed detonation model, WS15DD1 as a specific example here, shows the outermost density structure largely consistent with those of HV SNe and especially 1999aa-like SNe.  The typical density of NV SNe is smaller than both of these specific models.

However, we note that the direct comparison between the density structures derived by spectral modeling and specific explosion models may be misleading and potentially overstating the discrepancy. SNe Ia show diversity in their peak luminosities which is interpreted as a result of diversity in the production of $^{56}$Ni. Accordingly, most of the models have a capability to produce a variation in the $^{56}$Ni production, where controlling parameters are different for different model sequences (see Section \ref{sec:intro}). Omitting the details, the difference will translate to the ejecta kinematics through the energy generation, which can lead to the variation in the density structure for a given model sequence. Fig. \ref{fig:density_model} shows this exercise based on the WS15DD1 model, assuming that there is a variation of the kinetic energy by $\pm 50$\%\footnote{ This is not an extreme assumption. The mass of $^{56}$Ni has a variation of $\sim 0.4 - 1 M_\odot$ or even larger for NV and 1991T-like SNe \citep[e.g.,][for a review on the SN Ia diversity]{maeda2016}. If this would trace the variation in the nucleosynthesis products as a whole, the variation in the energy generation rate is more than a factor of two (and in principle can even be larger for the final kinetic energy as the binding energy of the WD is to be subtracted).} and that the ejecta structure responds to the difference in the kinetic energy in a self-similar manner. The model can largely cover the density structures of the sample of SNe Ia\footnote{ Note that the variation in the kinetic energy is considered here as a possible source of the variation within the same sequence; this would not create the `distinct' difference between the two sequences.}, where more energetic models correspond either to the 1991T/1999aa-like SNe or HV SNe, while less energetic models cover the density structure derived for NV SNe. Indeed, the explosion energy should be related to the amount of $^{56}$Ni produced in the explosion, in a way that a more energetic SN Ia has a larger amount of $^{56}$Ni and thus brighter. This expectation is indeed consistent with the idea that a more energetic explosion in the delayed-detonation may potentially be connected to the 1991T/1999a-like SNe and the less-energetic explosion within the same delayed-detonation framework may explain at least a fraction of the NV SNe; this possible link between the observed luminosity sequence and the delayed-detonation scenario has indeed been frequently discussed \citep{Mazzali2014,Zhang2016,Ashall2016}.

The amount of the unburnt carbon provides additional key constraint. The W7 model has a large amount of the unburnt carbon (0.032 $M_{\odot}$), while the WS15DD1 model has 0.00542 $M_{\odot}$. Apparently, the W7 model predicts the unburnt carbon mass substantially exceeding the values derived in this work,  indicating that the detonation should anyway be involved in SN Ia explosions.
The amount of the unburnt carbon predicted by the WS15DD1 model (or in general the delayed detonation model) is largely consistent with those found for the `carbon-rich NV--1991T/1999aa sequence'. Given that the delayed detonation model can also cover the density structure for the sample of SNe Ia including those in this carbon-rich sequence, it survives as a potential model for SNe Ia in the NV-1991T/1999aa sequence. 

In the delayed detonation model sequence of \citet{Iwamoto1999}, the unburnt C+O layer exist above $\sim$25,000 km s$^{-1}$. This overlaps with the velocity range for the 1999aa-like SNe where the unburn carbon is found (with the outermost velocity of $\gsim$30,000 km s$^{-1}$). A drawback in this interpretation/scenario is that the ejecta of the (carbon-rich) NV type, for which the unburnt C+O layer is found between $\sim$15,000 km s$^{-1}$ and $\sim$20,000 km s$^{-1}$, are found well below the prediction by the delayed detonation model.

An additional factor that can be a source of the observational diversity is the ejecta asymmetry and viewing angle effect. Given the stochastic nature of the deflagration trigger and the turbulent nature in the deflagration-flame propagation, the delayed detonation model should have some degree of the ejecta asymmettry \citep{Seitenzahl2013} and even global, one-sided asymmetry \citep{Maeda2010_2d_ddt,Maeda2011}. The density structure, and therefore the Si II velocity, can then be angle-dependent \citep{Maeda2010}. 

Another explosion scenario that deserves consideration is the double detonation model. Figure \ref{fig:density_model} shows the comparison between the density structures obtained in the present study and predicted by the double detonation model \citep{Shen2021}. In this model, the carbon core mass and the He shell mass are $\sim 0.98 M_{\odot}$ and $\sim 0.016 M_{\odot}$, respectively.

For the density structure, the double-detonation model is highly dependent on the viewing angle, and the explosion is generally `one-sided' in the most likely situation of the He detonation starting on a single point \citep{Fink2010}. We thus expect a high-density and high-velocity ejecta structure for one side while a low-density and low-velocity ejecta structure for the opposite direction. This behavior is seen in the model by \citet{Shen2021}, and a possible associate of the different viewing angles to different subclasses is indicative; the ejecta density in the `high-velocity' side ($\mu=+0.93$) roughly matches to the structure of either the 1991T/1999aa-like SNe Ia or the HV SNe Ia, while that in the `low-velocity' to `average-velocity' sides ($\mu=-0.93$ to $0$) does so to the NV SNe\footnote{ This is indeed a qualitatively similar argument based on an asymmetric delayed-detonation model to explain the HV and NV classes by the viewing angle effect \citep{Maeda2010}.}.

The unburnt carbon masses in the carbon core and the He shell are $0.0041 M_{\odot}$ and $0.00013 M_{\odot}$, respectively. The interface between the core and the shell is around the jump in the density structure. In Fig. \ref{fig:density_model}, it is seen that the velocity range as probed by the Tardis model results basically corresponds to the He shell. Therefore, in this particular model, the He shell is mostly observed with a minor contribution from the core materials. The amount of the unburnt carbon in the He shell is very small, which contradicts to that found for the SNe in the 1991T/1999aa-NV (C-rich and low-density) sequence. On the other hand, the model prediction on the carbon mass provides a qualitative match to the carbon-poor ejecta found for (most of) the SNe in the HV-NV (C-poor and high-density) sequence. 

As in the case for the delayed-detonation model (see above), the density structure should be dependent on the energy production (and the WD mass). Thus, this may provide an additional possibility to connect the HV-NV sequence in terms of the expected variation in the energy generation as mainly determined by the WD mass in this scenario. Additionally, the density in the He shell will also be affected by the amount of the He shell. Omitting the details and just assuming the variation of the kinetic energy by $\pm 50$\% with a fixed WD mass, we can roughly reproduce a range of the density structure covering both NV and HV SNe, considering that the variation in the viewing direction will introduce further diversity. Additional effects of the change in the progenitor WD mass and the He shell mass should further widen the expected range so that the range of the density structure derived with the  TARDIS models may easily be explained.  
However, this effect should not be a main driver to create the difference within the HV-NV sequence; little difference is seen in the luminosities of HV SNe and NV SNe, while the energy generation affects the $^{56}$Ni production. 

In summary, we might tentatively associate that the 1991T/1999aa--NV sequence and the HN--NV sequence might be linked to the delayed detonation mechanism and the double-detonation mechanism. The transition of the SN subclass in the 1991T/1999aa--NV sequence could be explained by the strength of the nuclear burning, forming the peak luminosity sequence. On the other hand, the diffence between the HV and NV SNe within the HV--NV sequence should not be explained by a similar effect, and the highly asymmetric ejecta and the viewing-angle effect may be a possible factor to form this sequence.  Note that there are also some problems that are not readily explained by the scenario provided here; for example, the outermost ejecta density predicted in the reference models used for the comparison is indeed higher for the delayed-detonation model than the double-detonation model, while the present study indicates the higher density for the HV--NV sequence than for the 1991T/1999aa--NV sequence.

\section{Discussion}\label{sec:discussion}
\subsection{Density at the photosphere}
\begin{figure}[t!]
\epsscale{1.2}
\plotone{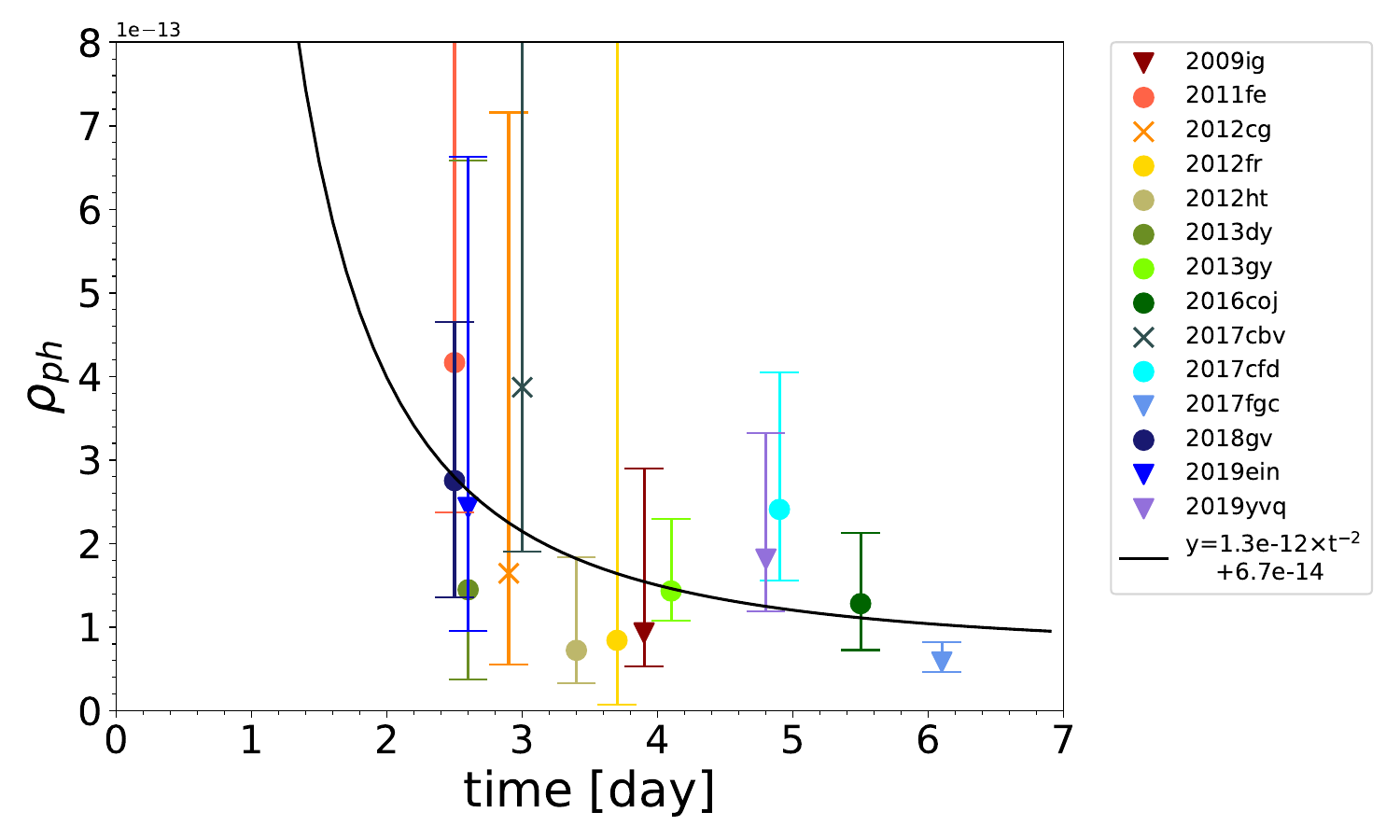}
\caption{The density at the photosphere for the sample of SNe Ia at different epochs. The black line is the curve described by $t^{-2}$. Different symbols are used for different subclasses (see the caption of Fig. \ref{fig:carbon_fraction}).
\label{fig:photospheric_density}}
\end{figure}

Figure \ref{fig:photospheric_density} shows the density at the photosphere, obtained for the best-fit models for the sample of spectra at different epochs. Overplotted by a black line is the curve described by $t^{-2}$, which roughly represents the evolution of the density at which the optical depth is unity  if the Rosseland Mean opacity is taken to be constant with time. Overall, the photospheric densities found for the early phase (within a week explosion) are clustered in the range between  $\sim6\times10^{-14} \rm{[g\ cm^{-3}]}$ and $\sim7 \times10^{-13} \rm{[g\ cm^{-3}]}$; while the errors for individual measurements can be large for some data points, the evolution is consistent with the $t^{-2}$curve. 

It  might indicate that the location of the photosphere depends on the density at that point, but is not sensitive to the temperature or the density outside it. This is expected for the following reason; in the very early phase, the density gradient is so steep that the continuum optical depth is mainly contributed by a small region just above the photosphere. What is not trivial is that the position of the photosphere is relatively insensitive to the temperature there, which is obtained as a result of the spectral formation computed by TARDIS as calibrated with the observed spectra; it indicates that a constant opacity is not a bad approximation to evaluate the position of the photosphere, despite a range of the photospheric temperature derived for different SNe Ia (sec.\ref{sec:temperature}).

\subsection{The photospheric temperature and the 'early-phase' Branch diagram}\label{sec:temperature}
 An established way to investigate the diversity in SNe Ia is the Branch diagram \citep{Branch2006}, specifically developed for the maximum-phase properties. The diagram provides a guide in studying the origin of the spectral diversity \citep[e.g.,][]{Nugent1995}. Now that we have extracted physical properties of the outermost ejecta, we are motivated to produce an analog of the Branch diagram but for the earliest phase and investigate the possible origin of the observed diversity. In addition, expanding the Branch diagram to various phases are interesting on its own; such investigation has been rare so far, especially for the earliest phase due to the paucity of the data. In this section, we first present the early-phase version of the Branch diagram (i.e., the spectral property), and investigate how it is related to the decline rate (i.e., the light-curve property) and then to the potospheric temperature (i.e., the outcome of the spectral modeling).

\begin{figure*}[t!]
\epsscale{0.45}
\hspace{-1cm}
\plotone{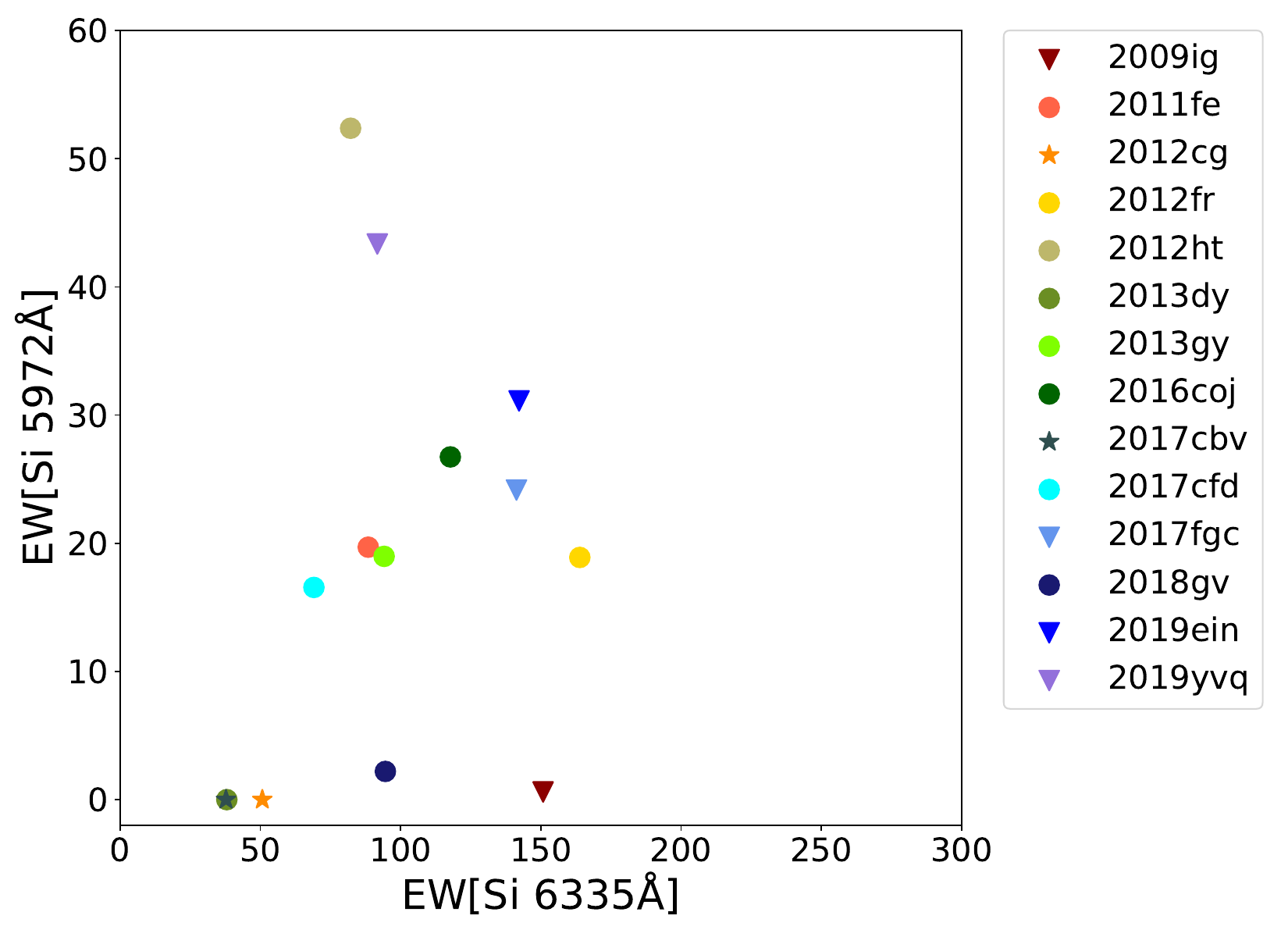}
\epsscale{0.38}
\hspace{-0.5cm}
\plotone{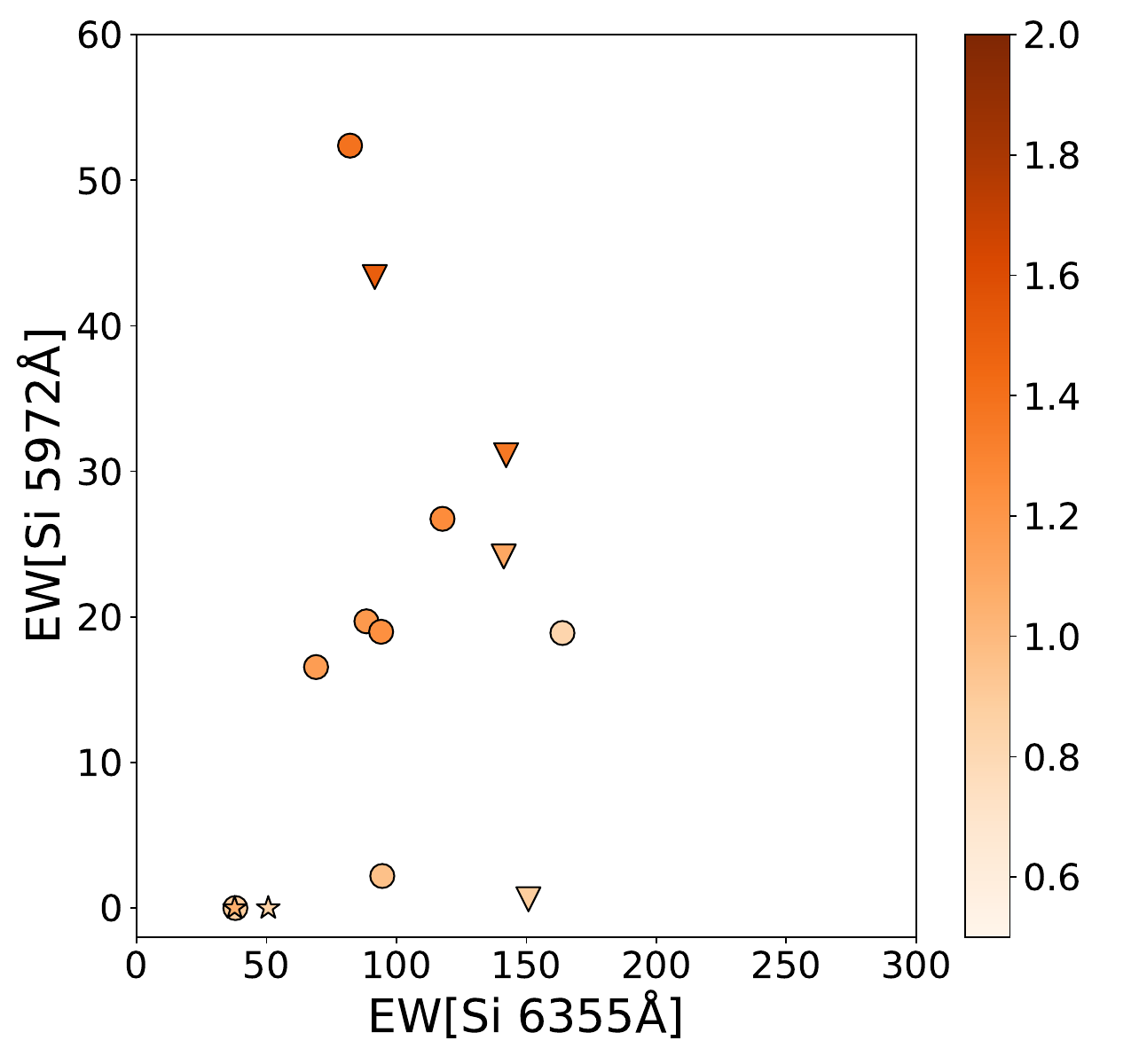}
\hspace{-0.5cm}
\plotone{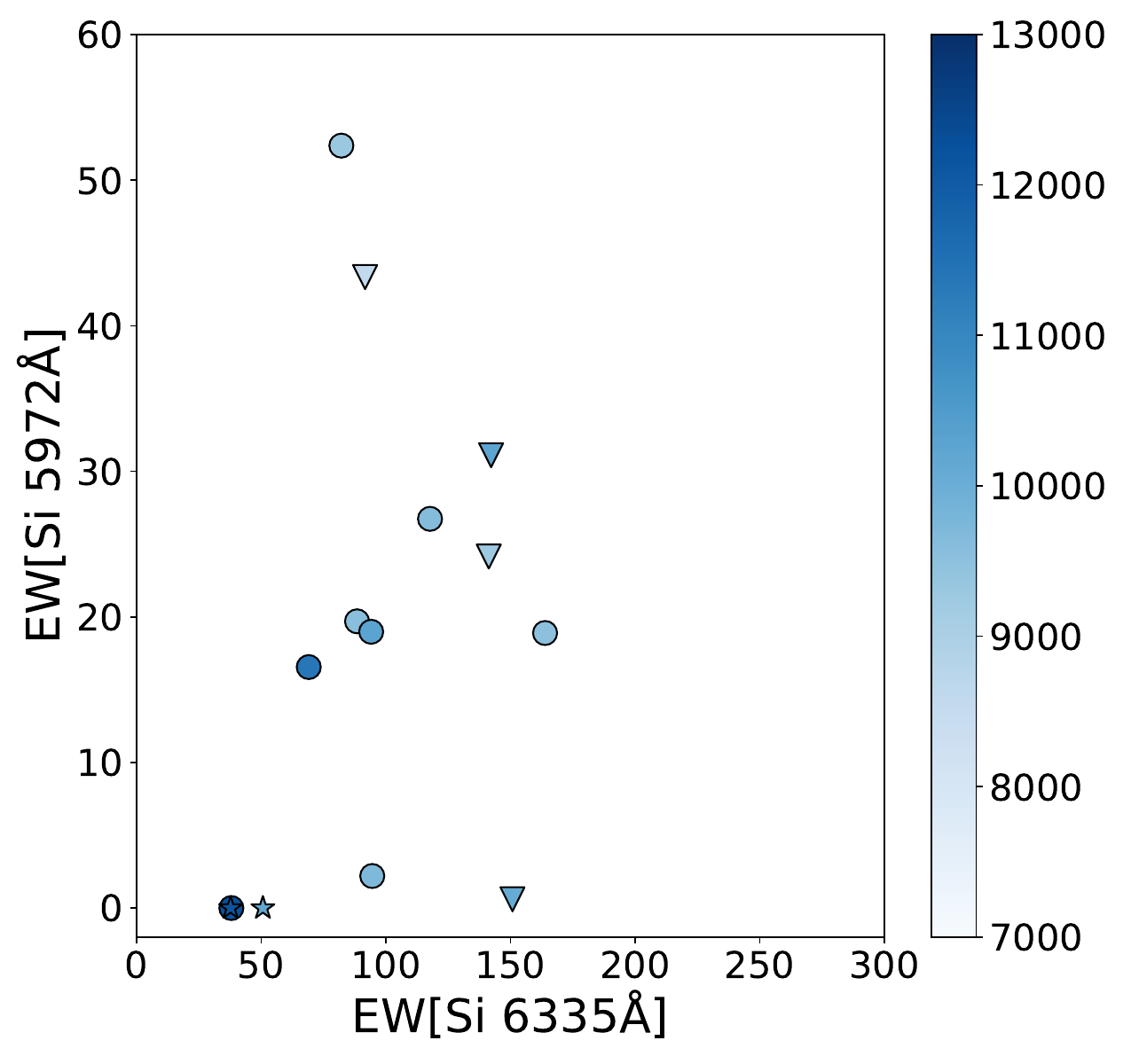}
\caption{ An analog of the Branch diagram for the very early-phase spectra we used (left). Circles represent the NV type, inverse triangles indicate the HV type, and stars correspond to the 1999aa-like type. The same figure but marked with $\Delta m_{15}$ is shown in the middle panel; color of the symbols indicate the value of $\Delta m_{15}$ following the color bar on the right side. The same but marked the photospehric temperature is shown in the right panel. 
}
\label{fig:Branch}
\end{figure*}

\subsubsection{Branch-like diagram using the very early-phase observables}
Figure \ref{fig:Branch} shows the early-phase version of the Branch diagram \citep{Branch2006}. In our sample we have two 1999aa-like SNe, which are classified as the `shallow-silicon' class based on their maximum-light spectra showing small EWs of both \ion{Si}{2} 5972 \rm{\AA} and \ion{Si}{2} 6355 \rm{\AA}. In the early-phase spectra (within a week since the explosion), they show the same behavior and can be distinguished from other subclasses in the same manner with the maximum-light classification scheme. Interestingly, the opposite is not the case; we see that SN 2013dy, which is classified as the `core-normal' class (which is nearly identical to the `NV' classification) in the maximum phase, does show the shallow-silicon characteristics in the early, rising phase. Namely, it should be categorized into the same group with the 1991T/1999aa-like subclass (or shallow-silicon class) in the very early phase, which is consistent with the result of the TARDIS modeling  (see Section \ref{sec:2013dy}). 

HV SNe are distinguished from other subclasses by their large EWs of \ion{Si}{2} 6355 \rm{\AA} in the maximum-phase Branch diagram. Fig. \ref{fig:Branch} shows that the same applies to their early-phase spectra. For HV SNe except 2019yvq, the EWs of \ion{Si}{2} 6355 \rm{\AA} are larger than 100\AA\ and those shown by the other subclasses. Similarly to the case for the 1991T/1999aa-like subclass, there is a transitional object which changes its spectral properties; SN 2012fr is in the NV (or core-normal) subclass based on the maximum-phase classification, but it clearly shares the spectral properties with HV SNe in the very early phase (see sec. \ref{sec:2012fr}). 

For most NV SNe, the EWs of \ion{Si}{2} 5972 \rm{\AA} and \ion{Si}{2} 6355 \rm{\AA} are around 20\AA\ and 80\AA, respectively. This is separated from the region occupied by the 1991T/1999aa-like and HV SNe, and thus they share their identity as the NV class even in the very early phase. As mentioned above, the opposite is not the case; we see a few example which turn their properties either from the 1991T/1999aa-like class or the HV class to the NV class toward the maximum phase, while in our sample we do not see clear examples of the opposite case, i.e., a transition from the NV class to the other classes. 

In summary, we conclude that the classification based on the maximum-light spectra generally holds in the very ealry-phase observable. Interestingly, a fraction of SNe change their classifications depending on the spectral phase (i.e., SNe 2012fr, SN 2013dy), from the 1991T/1999aa-like class or the HV class to the NV class (but not the opposite); this can be understood in view of their ejecta structures, as derived by the TARDIS modeling, bridging different subclasses in the outer and inner regions.

\subsubsection{Dependence on the declining rate}
In terms of $\Delta m_{15}$, the very early-phase Branch diagram may provide stronger diagnostics than in the maximum-light phase. 
Figure \ref{fig:Branch} shows $\Delta m_{15}$ distribution.
For HV SNe, there is a clear trend that SNe with larger $\Delta m_{15}$ are distributed in the upper region (i.e., a larger EW of \ion{Si}{2} 5972 \rm{\AA}). The same tendency can be seen for NV SNe. This behavior is especially clear for SN 2012ht and SN 2019yvq, whose $\Delta m_{15}$ is very large. They are distributed in the uppermost portion of the diagram, which may be analogous to the `cool' type classification in the maximum-light phase. While we do not have the cool type (or SN 1991bg-like SNe) in our sample, they might indeed represent a transitional case from the 1991bg-like (cool) subclass in the early phase to the NV (core-normal) subclass in the maximum-light phase. Adding a sample of 1991bg-like SNe in the rising phase is strongly encouraged to confirm this suggestion. For the shallow silicon type, $\Delta m_{15}$ tends to be small and they indeed occupy the lowest portion in this diagram.

\subsubsection{Dependence on the temperature}
\citet{Nugent1995} investigated the origin of the spectral classification, and showed that the photospheric temperature is a key quantity that controls the spectral properties along the luminosity sequence in the maximum-light phase. For high temperature, the EW of \ion{Si}{2} 6355\rm{\AA} is small and \ion{Si}{2} 5972\rm{\AA} line is rarely detectable. For the temperature between 8600 and 9800 K,  both of the \ion{Si}{2} 5972\rm{\AA} and \ion{Si}{2} 6355\rm{\AA} lines become strong. For the temperature under 8000 K, the EW of \ion{Si}{2} 5972\rm{\AA} is increasing to become comparable to or even stronger than the 6355\rm{\AA} line. 

Figure \ref{fig:Branch} shows the photospheric temperature distribution in the very early phase calculated by TARDIS on the (early-phase) Branch diagram. There is a clear trend between the temperature and the EWs of these lines, linking the SNe showing the higher EWs to the lower temperature. The trend is especially strong in the EW of \ion{Si}{2} 5972\rm{\AA}; this corresponds to the temperature sequence found by \citet{Nugent1995} for the maximum-light spectra. We note that the two SNe with the strongest \ion{Si}{2} 5972\rm{\AA} is indeed not the `cool type' in the maximum-light classification, but they might be classified as an analog of the cool type (1991bg-like) in the very early phase and transitions to other subclasses in the maximum light (see above). The HV SNe (or BL type) show relatively low temperature in the early rising phase, which is analogous to the low temperature derived for them based on the maximum-light spectra \citep{Tanaka2008,Hachinger2008}.

\subsection{rise time}
\begin{figure}[t!]
\epsscale{1.2}
\plotone{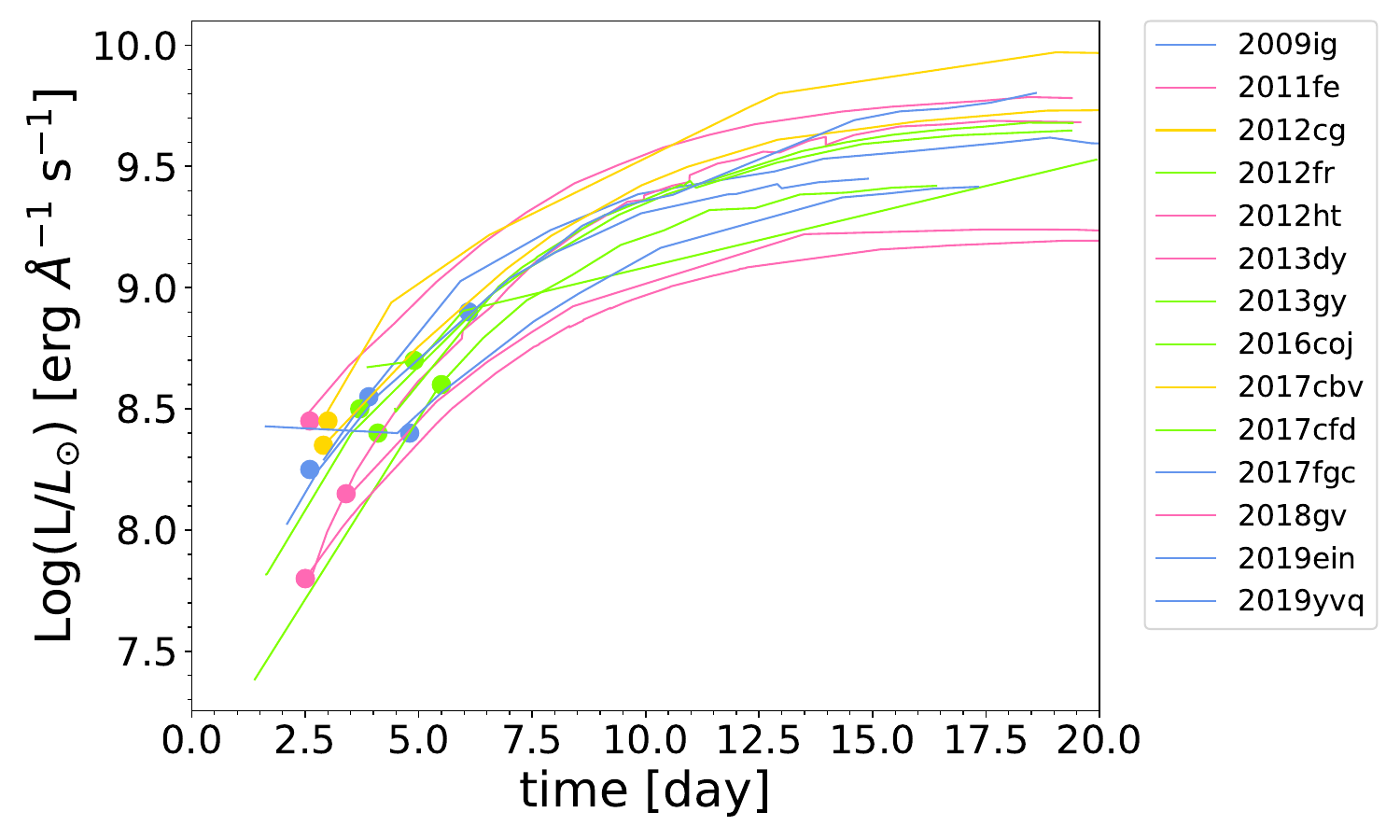}
\caption{Relation between time since the explosion and luminosity. The symbols represent the results of the TARDIS modeling. This is overlapped with the $R$- or $r$-band light curve for each object. Different colors are used for different subclasses (see the caption of Figure \ref{fig:density_model}). Note that SN 2011fe is hidden behind SN 2018gv. 
\label{fig:time_luminositu}}
\end{figure}

Figure \ref{fig:time_luminositu} shows the relation between time since the explosion and the luminosity, which are both obtained by the spectral fitting with TARDIS as shown by the symbols. These TARDIS points are used to anchor the $r$- or $R$-band light curves of the same objects shown by the lines with the same colors with the TARDIS points. There is a clear tendency that objects with lower $\Delta m_{15}$ show higher luminosity in the early rising phase. This is mainly explained by the difference in the peak luminosity, in a way that SNe that are bright in the maximum-phase tend to be bright also in the early phase. This is naturally expected from the relation between the peak luminosity and the peak date. 

SNe 1991T/1999aa-like SNe show small $\Delta m_{15}$ and thus high peak luminosity; they are thus on the top portion in the luminosity in the rising phase. No clear trend is seen about the rising luminosities and the other subclasses, i.e., the HV and NV SNe (in which latter may be further divided into the C-rich and C-poor) are not readily distinguishable, showing little trend beyond what the $\Delta m_{15}$ values would predict. The analysis here suggests two points in coordinating follow-up observations; (1) to study the early-phase `excess', one has to obtain highly-sampled multi-band light curves \citep{Burke2022}, as a single point which is calibrated by the spectral modeling does not help much to search for the early excess. (2) Oppositely, to extract the overall physical properties, a single spectrum can be more powerful than the high-cadence photometric observations.

\section{Conclusions}\label{sec:conclusion}
Aiming at constraining the explosion mechanisms, we have modeled very early-phase spectra of 14 SNe Ia. By using the one-dimensional Monte Carlo radiation transfer code, TARDIS, we have estimated the density structure and composition structure of the outermost ejecta. We have introduced  a semi-automatic fitting method to evaluate the quality of the fit between the synthesized and observed spectra, which  enhances the reproductivity as well as allows us to evaluate uncertainty in the derived parameters. 

We find that the outermost density structure is different for different subclasses. The HV type has a high density in the outermost layer, which extended to $\sim$30,000 km s$^{-1}$. The SN1999aa-like type shows the outermost ejecta extending to the similar velocity ($\sim$30,000 km s$^{-1}$), but with the density preferentially lower than in the HV class. The NV type does not show the high ejecta velocity as found in the other classes; they are mostly limited to $\sim$20,000 km s$^{-1}$. The densities of the ejecta between the photosphere ($\sim$15,000 km s$^{-1}$) and the outermost layer ($\sim$20,000 km s$^{-1}$) of the NV SNe show a substantial diversity by nearly an order of magnitude. 

We have also shown that the composition structures are divided into two groups; those dominated by unburnt C+O-rich layer and the others dominated by the carbon-burning layer with little trace of the unburnt C+O layer. Interestingly, the 1991T/1999a-like SNe and the HV SNe are clearly divided in this composition structure; C-rich for the 1991T/19991aa-like class and C-poor for the HV class. The NV class shows diversity; some show the C-rich composition while the other show the C-poor composition. There is a tendency that the C-rich NV SNe tend to show lower density than the C-poor NV SNe. 

According to these investigations, we suggest that the NV SNe may indeed be divided into two classes, one connected to the 1991T/1999aa-like class and the other to the HV class. Namely, the whole SN Ia sample analyzed here may be categorized into two populations; one with C-rich and low-density outermost layer (the 1991T/1999aa-like class and some NV SNe) and the other with C-poor and high-density outermost layer (the HV class and other NV SNe). Accordingly, the SNe may be divided into two sequence. 

These two sequences are likely associated with different origins in their progenitors and explosions. As one possibility, we have discussed how the two popular scenarios, the delayed-detonation mechanism on a $M_{\rm Ch}$ WD and the double-detonation on a sub-$M_{\rm Ch}$ WD, could explain the properties of the ejecta as derived by the TARDIS modeling for different sequence and subclasses. One possibility is this; the delayed-detonation mechanism is associated with the 1991T/1999aa-NV (C-rich and low-density) sequence, where the strength of the burning may control a transition between 1991T/1999aa-like SNe and NV SNe; the double-detonation mechanism is associated with the HV-NV (C-poor and high-density) sequence, where the viewing angle may be a driving factor to create the transition between HV and NV SNe. 

We have also plotted the early-phase spectral properties in the very early phase, in the same manner with the Branch diagram established for the maximum-light properties. Different subclasses form different sequences in this diagram, similar to what is found for the maximum-phase properties. As in the maximum-phase phase, the different subclasses can be characterized mainly by the photospheric temperature, while the HV class shows a relatively large diversity in the temperature. Interestingly, there is one NV SN that have similar properties with the 1991T/1999aa-like class in the early-phase Branch diagram, one NV SN that is indistinguishable with the HV SNe. These `transitions' are interpreted that they have outermost ejecta properties more closely related to other subclasses than the NV class. Namely, the outermost ejecta of the NV class show a larger diversity than in the inner ejecta. Furthermore, the NV SN (2013dy) which shows the transition from the 1991T/1999-like to the NV class is indeed in the C-rich, low-density sequence (i.e., the 1991T/1999aa--NV sequence), and the NV SN (2012fr) which shows the transition from the HV class to the NV class is in the C-poor, high-density sequence (i.e., the HV--NV sequence); this strengthens our argument for the existence of the the sequences. Interestingly, we have not identified the opposite transition, i.e., the NV class to either the 1991T/1999a-like class or the HV class. 
In addition, the early-phase Branch diagram suggests that one HV and one NV might be considered as an analog of the `cool' (1991bg-like) class in the maximum-light phase, further pointing to the same picture.

\begin{acknowledgments}

The authors thank Daichi Hiramatsu, Jian Jiang, Masaomi Tanaka, and Kohki Uno for stimulating discussion. K.M. acknowledges support from the Japan Society for the Promotion of Science (JSPS) KAKENHI grant JP18H05223, JP20H00174.  Numerical computation in this work was carried out at the Yukawa Institute Computer Facility.
Some data presented in this work are obtained from WISeREP (https://www.wiserep.org). This research made use of TARDIS, a community-developed software package for spectral synthesis in supernovae. The development of TARDIS received support from GitHub, the Google Summer of Code initiative, and from ESA's Summer of Code in Space program. TARDIS is a fiscally sponsored project of NumFOCUS. TARDIS makes extensive use of Astropy and Pyne.
\end{acknowledgments}

\software{$\texttt{TARDIS}$ v2022.06.19 \citep{Kerzendorf2014}}

\bibliography{main}{}
\bibliographystyle{aasjournal}

\end{document}